\documentclass[]{agujournal2019}
\usepackage{url} %this package should fix any errors with URLs in refs.  
\usepackage{lineno}
\usepackage[inline]{trackchanges} %for better track changes. finalnew option will compile document with changes incorporated.
\usepackage{soul}
\usepackage{array}
\usepackage{cellspace}
\usepackage{multirow}
\usepackage[squaren, Gray, cdot]{SIunits}
\usepackage{amsmath}

\usepackage{lmodern}

\usepackage{fancyhdr}

\usepackage{amsthm}
\usepackage{amsmath}
\usepackage{amssymb}
\usepackage{mathrsfs}

\usepackage{graphicx}
\graphicspath{{Figures_intro/}{Figures_Chap1/}{Figures_Chap2/}{Figures_Chap3/}{Figures_Chap4/}{Figures_Chap5/}{Figures_Annexes/}{Figures_page_garde/}}

\usepackage{lettrine}

\usepackage{enumerate}

\usepackage{cellspace}

\usepackage{array}

\usepackage{slashbox}

\usepackage{multirow}

\usepackage{placeins}

\usepackage{float}

\usepackage{footnote}
\makesavenoteenv{equation}
\makesavenoteenv{tabular}

\usepackage{titlesec}

\usepackage{textcomp}

\usepackage{multicol}

\usepackage{wrapfig}

\usepackage{bibentry}

\draftfalse

\journalname{}

\begin{document}

%% ------------------------------------------------------------------------ %%
%  Title
%
% (A title should be specific, informative, and brief. Use
% abbreviations only if they are defined in the abstract. Titles that
% start with general keywords then specific terms are optimized in
% searches)
%
%% ------------------------------------------------------------------------ %%

% Example: \title{This is a test title}

\title{Airborne absolute gravimetry with a quantum sensor, comparison with classical technologies}

%% ------------------------------------------------------------------------ %%
%
%  AUTHORS AND AFFILIATIONS
%
%% ------------------------------------------------------------------------ %%

% Authors are individuals who have significantly contributed to the
% research and preparation of the article. Group authors are allowed, if
% each author in the group is separately identified in an appendix.)

% List authors by first name or initial followed by last name and
% separated by commas. Use \affil{} to number affiliations, and
% \thanks{} for author notes.
% Additional author notes should be indicated with \thanks{} (for
% example, for current addresses).

% Example: \authors{A. B. Author\affil{1}\thanks{Current address, Antartica}, B. C. Author\affil{2,3}, and D. E.
% Author\affil{3,4}\thanks{Also funded by Monsanto.}}

\authors{Y. Bidel\affil{1}, N. Zahzam\affil{1}, A. Bresson\affil{1}, C. Blanchard\affil{1}, A. Bonnin\affil{1}, J. Bernard\affil{2}, M. Cadoret\affil{2},T.E. Jensen\affil{3}, R. Forsberg\affil{3}, C. Salaun\affil{4}, S. Lucas\affil{4}, M.F. Lequentrec-Lalancette \affil{4}, D. Rouxel\affil{4}, G. Gabalda\affil{5}, L. Seoane\affil{5}, D.T. Vu\affil{5}, S. Bonvalot\affil{5}}

 \affiliation{1}{DPHY, ONERA, Université Paris Saclay, 91123 Palaiseau,
France}
\affiliation{2}{LCM-CNAM, 61 Rue du Landy, 93210 La Plaine
Saint-Denis, France}
\affiliation{3}{National Space Institute, Technical University of Denmark, Elektrovej 327, 2800 Lyngby, Denmark}
\affiliation{4}{Shom – French hydrographic and oceanographic office, CS 92803, 29228 Brest, France}
\affiliation{5}{GET (CNRS, IRD, UPS, CNES) - University of Toulouse, Toulouse, France}

%(repeat as many times as is necessary)

%% Corresponding Author:
% Corresponding author mailing address and e-mail address:

% (include name and email addresses of the corresponding author.  More
% than one corresponding author is allowed in this LaTeX file and for
% publication; but only one corresponding author is allowed in our
% editorial system.)

% Example: \correspondingauthor{First and Last Name}{email@address.edu}

\correspondingauthor{Yannick Bidel}{yannick.bidel@onera.fr}

%% Keypoints, final entry on title page.

%  List up to three key points (at least one is required)
%  Key Points summarize the main points and conclusions of the article
%  Each must be 140 characters or fewer with no special characters or punctuation and must be complete sentences

% Example:
% \begin{keypoints}
% \item	List up to three key points (at least one is required)
% \item	Key Points summarize the main points and conclusions of the article
% \item	Each must be 140 characters or fewer with no special characters or punctuation and must be complete sentences
% \end{keypoints}

\begin{keypoints}
\item Demonstration of airborne absolute gravity measurements with a quantum sensor
\item Estimated measurement errors ranging from 0.6 to 1.3 mGal 
\item Precision equal or better than classical gravimeters
\item Good agreement with marine, land and altimetry derived gravity data
\item High potential for mapping shallow water or mountainous areas
\item Interest for linking ground and satellite measurements with homogeneous absolute referencing
\end{keypoints}

%% ------------------------------------------------------------------------ %%
%
%  ABSTRACT and PLAIN LANGUAGE SUMMARY
%
% A good Abstract will begin with a short description of the problem
% being addressed, briefly describe the new data or analyses, then
% briefly states the main conclusion(s) and how they are supported and
% uncertainties.

% The Plain Language Summary should be written for a broad audience,
% including journalists and the science-interested public, that will not have 
% a background in your field.
%
% A Plain Language Summary is required in GRL, JGR: Planets, JGR: Biogeosciences,
% JGR: Oceans, G-Cubed, Reviews of Geophysics, and JAMES.
% see http://sharingscience.agu.org/creating-plain-language-summary/)
%
%% ------------------------------------------------------------------------ %%

%% \begin{abstract} starts the second page

\begin{abstract}

We report an airborne gravity survey with an absolute gravimeter based on atom interferometry and two relative gravimeters: a classical LaCoste\&Romberg (L\&R) and a novel iMAR strap-down Inertial Measurement Unit (IMU). We estimated measurement errors for the quantum gravimeter ranging from 0.6 to 1.3 mGal depending on the flight conditions and the filtering used. Similar measurement errors are obtained with iMAR strapdown gravimeter but the long term stability is five times worse. The traditional L\&R platform gravimeter shows larger measurement errors (3 - 4 mGal). Airborne measurements have been compared to marine, land and altimetry derived gravity data. We obtain a good agreement for the quantum gravimeter with standard deviations and means on differences below or equal to 2 mGal. This study confirms the potential of quantum technology for absolute airborne gravimetry which is particularly interesting for mapping shallow water or mountainous areas and for linking ground and satellite measurements with homogeneous absolute referencing.

\end{abstract}

\section*{Plain Language Summary}

Quantum technology offers a new kind of sensor for airborne gravimetry. Contrary to classical technologies which can only measure variation of gravity from an aircraft, a quantum gravimeter provides directly an absolute measurement of gravity eliminating the necessity of calibrations and drift estimations. We report here an airborne survey with a quantum gravimeter and two classical gravimeters. We demonstrated that the quantum gravimeter reaches the same precision than the best classical gravimeter. The gravity measurements have also been validated with models derived from land and marine gravity measurements and satellite altimetry.

%% ------------------------------------------------------------------------ %%
%
%  TEXT
%
%% ------------------------------------------------------------------------ %%

%%% Suggested section heads:
% \section{Introduction}
%
% The main text should start with an introduction. Except for short
% manuscripts (such as comments and replies), the text should be divided
% into sections, each with its own heading.

% Headings should be sentence fragments and do not begin with a
% lowercase letter or number. Examples of good headings are:

% \section{Materials and Methods}
% Here is text on Materials and Methods.
%
% \subsection{A descriptive heading about methods}
% More about Methods.
%
% \section{Data} (Or section title might be a descriptive heading about data)
%
% \section{Results} (Or section title might be a descriptive heading about the
% results)
%
% \section{Conclusions}

\section{Introduction}

A new technology of gravimetry based on atom interferometry \cite{Berman1997} is emerging. It is particularly promising because it confers at the same time absolute measurements, long term stability, high sensitivity and robustness. No classical instruments regroup all these advantages. Indeed, quantum gravimeters can have the same accuracy as falling corner cube gravity instruments \cite{Karcher2018}. Like superconducting gravimeters, it is used to continuously monitor gravity with high long term stability \cite{Freier2016, Menoret2018}. It has been also demonstrated that such technology could be implemented on moving vehicles like spring gravimeters or forced balanced accelerometers \cite{Bidel2018, Bidel2020, Huang2022}. Atom interferometry technology is finally also studied for the next generation of sensor for space gravimetry \cite{Zahzam2022, Leveque2021, Reguzzoni2021, Trimeche2019}. 

In this paper, we focus on airborne gravimetry \cite{Forsberg2010} which is a powerful tool for regional gravity mapping. It allows higher spatial resolution than space gravimetry \cite{Sandwell2014, Kvas2019, Pail2011} and can cover areas which are difficult to map with ground gravimeters like mountains, glaciers or deserts. Airborne gravimetry is also especially interesting in the coastal areas where satellite altimetry is not precise. In this context, quantum gravimeters are particularly interesting for airborne surveys because it is the only technology that provides absolute gravity measurements. Other technologies as spring gravimeters or force balanced accelerometers provide only variation of gravity and suffer from instrumental drift. Quantum gravimetry could thus make airborne surveys faster, cheaper and more precise since there is no need for calibration and drift estimations. The technology might also be very useful for correcting
historical land and marine gravity data, which are often biased.

Within the last few years, a dynamic quantum sensor developed by the French aerospace lab ONERA for the hydrographic and oceanographic marine office (Shom) and the French Defence Agency (DGA) has been successfully tested in field conditions. Results derived from shipborne surveys and comparisons with other conventional marine gravity meter (Bodensee KSS32) and with satellite altimetry data confirmed the ability of quantum technologies for measuring absolute gravity in dynamic surveys with unprecedented robustness and accuracy below 1 mGal \cite{Bidel2018}. Initially designed for marine surveys, the same quantum sensor (hereafter called GIRAFE) has been also tested during an airborne survey. This first experiment also confirmed the capabilities of such quantum technologies for measuring absolute gravity from an aircraft with precision measurements ranging from 1.7 to 3.9 mGal \cite{Bidel2020}. Considering the promising perspectives derived from these results for future geophysical surveys, a dedicated airborne campaign has been proposed for assessing the potentialities and accuracy of airborne absolute gravimetry for surveying coastal, marine and mountainous areas \cite{Bonvalot2018}. An improved version of the GIRAFE quantum gravimeter used in the previous campaign was used and compared with a LaCoste\&Romberg (L\&R) platform gravimeter and with an iMAR strapdown gravimeter. Here we report the results from this airborne gravity surveys carried out in selected areas where data acquisition still remains challenging (land-sea transition and mountain range) and from a reference profile. Quantum gravity data measurements are compared with simultaneous measurements performed by other conventional instruments and with available information coming from satellite, marine and terrestrial gravity measurements.

This article is organized as follows. In the first part, the quantum gravimeter and the two relative gravimeters are shortly described insisting on the modification of the quantum gravimeter compared to the last campaign \cite{Bidel2020}. Then, in the second part, airborne data acquisitions are described and computed gravity disturbances are shown. Ground gravity measurements acquired during the campaign are also presented to check the accuracy and the long term stability of GIRAFE gravimeter. In the third part, the errors of airborne measurements are estimated from repeated measurements at the same location. In the fourth part, gravity measured by the three instruments is inter-compared. Finally, in the last part, airborne gravity measurements are compared to models derived from land, marine gravity measurements and satellite altimetry.

\section{Dynamic gravity meters}

\subsection{GIRAFE quantum gravimeter}

The quantum gravimeter (GIRAFE) used during this campaign is an improved version of the quantum gravimeter previously tested on a boat \cite{Bidel2018} and aircraft \cite{Bidel2020}. Details of the principles and main characteristics of the instrument can be found in the above mentioned papers. In this part, we will present a brief description of the gravimeter, emphasizing the improvements made compared to the previous airborne campaign \cite{Bidel2020}. 

The gravimeter is composed of an absolute atom accelerometer, a gyro-stabilized platform which maintains the accelerometer aligned with the local gravity vector and systems which provide the lasers and microwaves needed to the atom sensor and perform data acquisition and processing. 

The absolute accelerometer is based on the acceleration measurement of a free falling gas of cold atoms by atom interferometry \cite{Berman1997,Tino2014}. The measurement sequence has three steps. First, a cloud of cold atoms is prepared by laser cooling and trapping method \cite{Metcalf2007}. Then, the free falling atoms are submitted to three laser pulses which split, redirect and recombine the atom wave function. Finally, the signal of atom interference is recorded by fluorescence detection. The signal obtained is proportional to the cosine of the acceleration $a$ :
\begin{equation}
P=P_m+\frac{C}{2} \cos\left(\frac{4\pi}{\lambda}\times a \times T^2\right)
\end{equation}
In this expression, $\lambda=780$ nm is the wavelength of the laser performing atom interferometry and $T$ is the duration between the laser pulses which is equal to 20 ms in nominal condition or 10 ms in presence of large variations of acceleration. Our sensor provides measurements at a repetition rate of 10 Hz. The atom accelerometer has the advantage to provide an absolute measurement of the acceleration but it has measurement dead times during the cold atom preparation and the detection. Moreover, many values of acceleration are possible for a given signal of the atom sensor due to the inversion of the cosine function. To overcome these limitations, the atom accelerometer is hybridized to a conventional accelerometer which fills the measurement dead times and lifts the ambiguity of the cosine function by providing an approximate value of the acceleration. On the other hand, the atom accelerometer estimates continuously the bias of the classical accelerometer.

In this improved version of the gravimeter, we modified the hybridization protocol by estimating continuously also the scale factor of the auxiliary conventional accelerometer in GIRAFE. The control loop for the correction of the scale factor is using an error signal equal to $\Delta a(a-g_0)$ where $\Delta a$ is the acceleration difference measured between the classical and the atom accelerometer and $g_0=9.8$ m s$^{-2}$ is the mean gravity. This improvement is particularly important during turbulent parts of a flight where there are large variations of acceleration. The horizontal lever arm between the measurement points of the the atom and the classical accelerometer has been decreased. This horizontal lever arm is now 0.4 mm instead of 4.2 mm previously. With this improvement, we obtain a better agreement between the accelerations measured by the atom and the classical sensor in presence of angular accelerations. We also synchronize the measurement sequence with a Pulse Per Second (PPS) signal coming from  the Global Navigation Satellite System (GNSS). This allows to have a perfect synchronization between the measurement of the gravimeter and GNSS navigation data and thus improving the correction of kinematic acceleration estimated by GNSS. Missing measurement points \cite{Bidel2020} have now been almost completely eliminated. Finally, the laser system has been improved by using now an all fiber laser system that is not subjected to possible misalignments. Indeed, in our laser system based on a frequency doubled telecom fiber bench \cite{Carraz2009}, the free space frequency doubling in a bulk crystal has been replaced by a frequency doubling in a waveguide crystal \cite{Leveque2014}. 

In static conditions and for $T=20$ ms, the measurement sensitivity of the gravimeter is equal to 0.8 mGal/Hz$^{1/2}$ and the accuracy is equal to 0.17 mGal \cite{Bidel2018}. 

\subsection{Classical gravimeters}

Two conventional airborne gravimeters were additionally installed on board the aircraft. The first was an older L{\&}R S-type gravimeter mounted on a two-axis damped platform \cite{1992_Valliant}, the second a navigation-grade IMU with temperature stabilization from iMAR Navigation, mounted in a strapdown configuration \cite{2019_Jensen_et_al}. Both of these instruments represent relative gravimeters, measuring variations in the gravity field with respect to base readings performed on the ground before and after flight. The strapdown IMU system contains an internal GNSS receiver providing time-stamped observations synchronized with respect to GNSS navigation data. The L{\&}R observations are time synchronized by deriving vertical accelerations from an external GNSS receiver and applying a constant time shift to optimize correlation between both acceleration data streams.

%-----------------------------------------------------------------------------------------------------------------------------------
\FloatBarrier

\section{Airborne gravity data acquisition and processing}
\subsection{Main description and objectives}

The airborne gravity campaign took place in April - May 2019 across France, using an ATR-42 aircraft from SAFIRE (French facility for airborne research) located in Toulouse Francazal airport. 

As mentioned above, the main objectives were (1) to assess the accuracy of the quantum GIRAFE instrument in comparison with other conventional gravity meters currently used for airborne gravimetry and (2) to evaluate the added value of such instrumentation for improving the gravity mapping of areas poorly covered by surface or satellite gravity measurements all such as land-sea transitions and mountainous regions. To fulfil these objectives, we first installed on board the aircraft two classical airborne gravimeters for performing simultaneous data acquisition with the quantum GIRAFE instrument.  Secondly, two survey areas were selected on the western Atlantic shore over Bay of Biscay where both deep (up to -2000m) and shallow waters could be surveyed, and a southern area encompassing progressively the Pyrenees mountain range (heights up to 3000m or more) and flat regions (heights from 150 - 200 m). In addition, a reference profile located off shore of Brittany and currently used by Shom for testing marine gravimeters has been repeated for assessing and comparing the accuracy of all embarked instruments. This reference line, also measured with GIRAFE instrument during the previous shipborne surveys \cite{Bidel2018} will also enable to compare the performances of GIRAFE quantum meter in both airborne and shipborne surveying. 

\begin{figure}[h!]
\vspace{0cm}
 \noindent\includegraphics[width=12 cm,natwidth=1121 px,natheight=1633px]{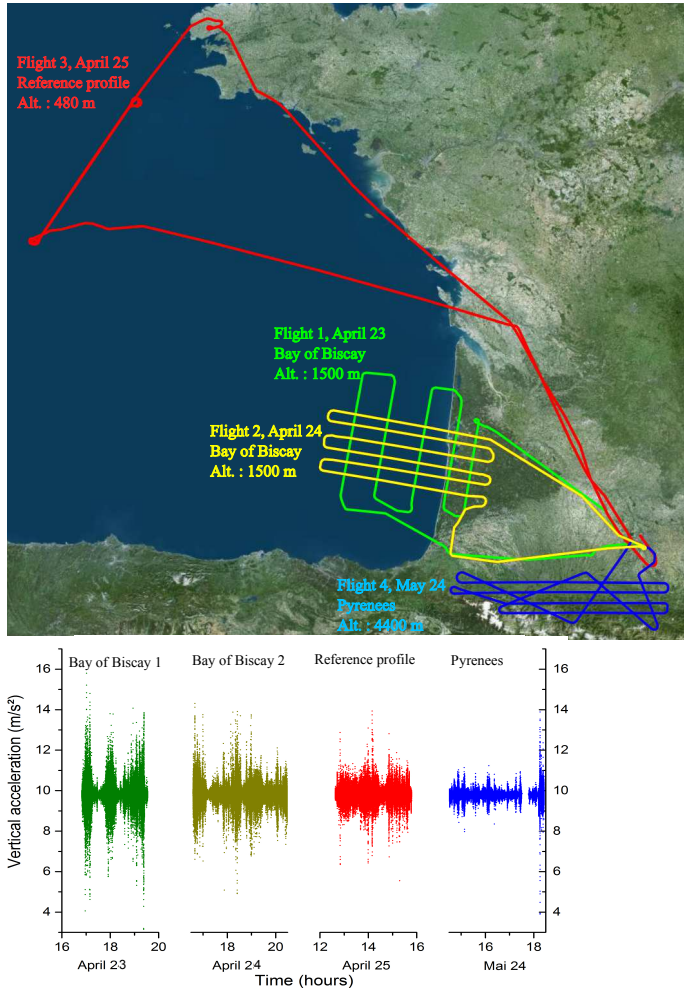}
\caption{Top: Flight plan of the gravity campaign. Bottom:  Vertical accelerations measured by the quantum gravimeter at a rate of 10 Hz during the flights.}
\label{fig_survey}
\end{figure}

The flight lines for the three selected areas surveyed during spring 2019 are given in Figure \ref{fig_survey} and Table \ref{Flights}. They include four gravity measurements ﬂights (approximately 30 hours) acquired as follows. The first and second flights took place on April 23 and 24 and consisted in a survey over the Bay of Biscay in order to demonstrate the ability to map the land sea transition. The first flight was dedicated to measure gravity along 5 North-South lines and the second flight was dedicated to measure gravity along 6 West-East lines. During these two flights, the altitude was 1500 m. The third flight took place on April 25 and consists of five repeated measurements over the reference profile off shore of Brest at an altitude of 480 m.  Finally, the fourth flight took place on May 24 over the Pyrenees for demonstrating the ability to map mountain areas. We measured gravity above Pyrenees along 9 lines at an altitude of 4400 m. 

The along-track plane velocity was around $v=100$ m/s. The vertical accelerations experienced by the gravimeters during the four gravity measurement flights are given in Figure \ref{fig_survey}. During turbulent parts, acceleration variations can reach 10 m/s² peak-peak and during quiet parts acceleration variation are below 1 m/s² peak-peak.

\begin{table}[h!]
 \caption{ Overview of the measurement flights and characteristics.}
 \centering
 \begin{tabular}{l l l l l }
Flight \#				&	Date						& Survey						& Altitude & Descriptions		\\
\hline
1								& April 23, 2019	& Bay of Biscay			& 1500 m	 &		5 North-South lines	\\
\hline
2							  & April 24, 2019	& Bay of Biscay			& 1500 m	 &	  6 West-East lines	\\
\hline	
3								& April 25, 2019	& Reference	profile & 480 m		 &		5 repeated lines		\\ 
\hline	
4								& May 24, 2019	 & Pyrenees						& 4400 m	 &	  9 lines			\\ 								
 \end{tabular}
\label{Flights}
 \end{table}

\FloatBarrier 

\subsection{Ground measurements}

\begin{figure}[h!]
 \noindent\includegraphics[width=20 cm,natwidth=2491px,natheight=789px]{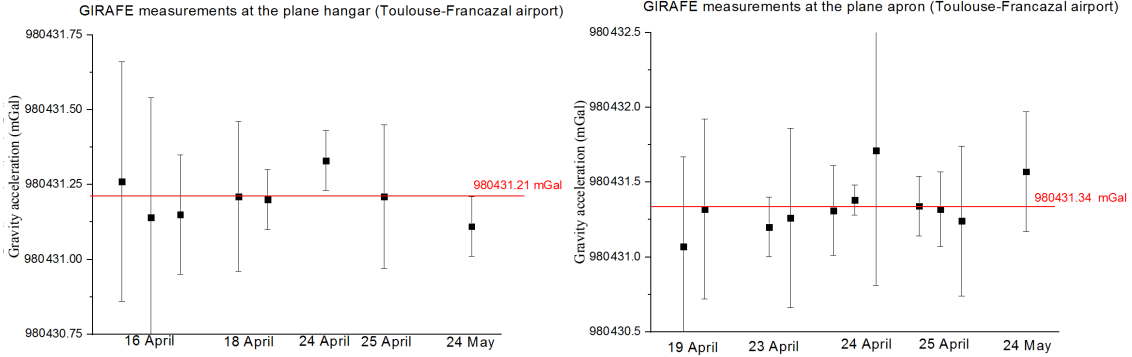}
\caption{Ground gravity measurements with GIRAFE gravimeter installed in the aircraft. These measurements were done in Toulouse Francazal airport in the SAFIRE hangar (left) and on the apron (right). The error bars represent the statistical uncertainty of measurements and depend  on the amplitude of the aircraft movement and on the duration of the measurements which are ranging from 8  to 60 min. The red line is the weighted average of the displayed gravity measurements.}
\label{fig_GroundMeas}
\end{figure}

Throughout the airborne campaign from 19 April to 24 May, ground gravity measurements in the plane hangar and in the apron were performed by GIRAFE gravimeter installed in the aircraft. The results are given in Figure \ref{fig_GroundMeas}.  The set scatter of absolute gravity values is 0.22 mGal peak-peak for measurements in the plane hangar and 0.64 mGal peak-peak for measurements in the plane apron. These set scatters agree with the estimated statistical uncertainty of the measurements confirming the long term stability of the gravimeter even after the shocks, vibrations and accelerations which occur during flight and after night electrical shut down. 

In order to provide an accurate ground reference gravity value for the airborne survey, absolute gravity measurements were also carried out during the campaign at both Toulouse-Francazal and Brest airports using A10 Micro-g LaCoste absolute gravity meters, A10 \#014 and A10 \#031 respectively. The absolute gravity values determined on ground within few tens of meters from the aircraft with the A10 instruments, were tied at the mean location of the instrument using CG-6 relative gravity meters. A first series of measurements was carried out in the hangar of Toulouse Francazal airport where the aircraft was parked. A second series was done on the apron of Toulouse Francazal airport where the aircraft was parked just before the take off. Finally, gravity measurements were done in Brest airport tarmac on the occasion of a refuelling for the flight off shore over the reference profile. The gravity values obtained by CG-6 and GIRAFE gravimeters are shown in Table \ref{ground_meas}. For Toulouse Francazal and Brest airport aprons, the measurements of GIRAFE gravimeter are in excellent agreement  with the measurements of the calibrated CG-6 ($\leq$ 0.1 mGal), while for Toulouse Francazal airport hangar, we obtain a difference of 0.3 - 0.4 mGal, larger that the estimated uncertainty which we could not explain.

\begin{table}[h!]
\scriptsize
 \caption{Gravity ground measurements. Tidal corrections \cite{Tamura1987} were applied to the CG-6 and GIRAFE gravimeters.}
 \centering
 \begin{tabular}{l l l l l }
Location										&	Gravimeter							& Gravity measurement	& Height			& Difference GIRAFE/CG-6 with \\
														&													&			(mGal)					&		(m)				& height correction (0.30 mGal/m)\\
														&													&											&							& (mGal) \\
 \hline
Toulouse Francazal 					& GIRAFE on the ground		& 980 431.50  $\pm$ 0.17	& 1.03		& 0.36 $\pm$ 0.17\\
airport hangar							& GIRAFE in the aircraft	& 980 431.21  $\pm$ 0.18	& 2.1 		& 0.39 $\pm$ 0.18\\
														& CG-6 on the ground			& 980 431.453 $\pm$ 0.005	& 0			  &\\
\hline
Toulouse Francazal 				  & GIRAFE in the aircraft	& 980 431.34  $\pm$ 0.18	& 2.1 		& 0.11 $\pm$ 0.18\\
airport apron								& CG-6 on the ground			& 980 431.861 $\pm$ 0.009 & 0 			&\\
\hline	
Brest airport 						  & GIRAFE in the aircraft	& 980 915.08  $\pm$ 0.30 	& 2.1 		& 0.06 $\pm$ 0.3\\
apron 											& CG-6	on the ground			& 980 915.646 $\pm$ 0.008	& 0     	&\\													
 \end{tabular}
\label{ground_meas}
 \end{table}

\FloatBarrier 

\subsection{Computation of gravity disturbances}

\begin{figure}[h!]
\vspace{-2cm}
 \noindent\includegraphics[width=19 cm,natwidth=3100px,natheight=2145px]{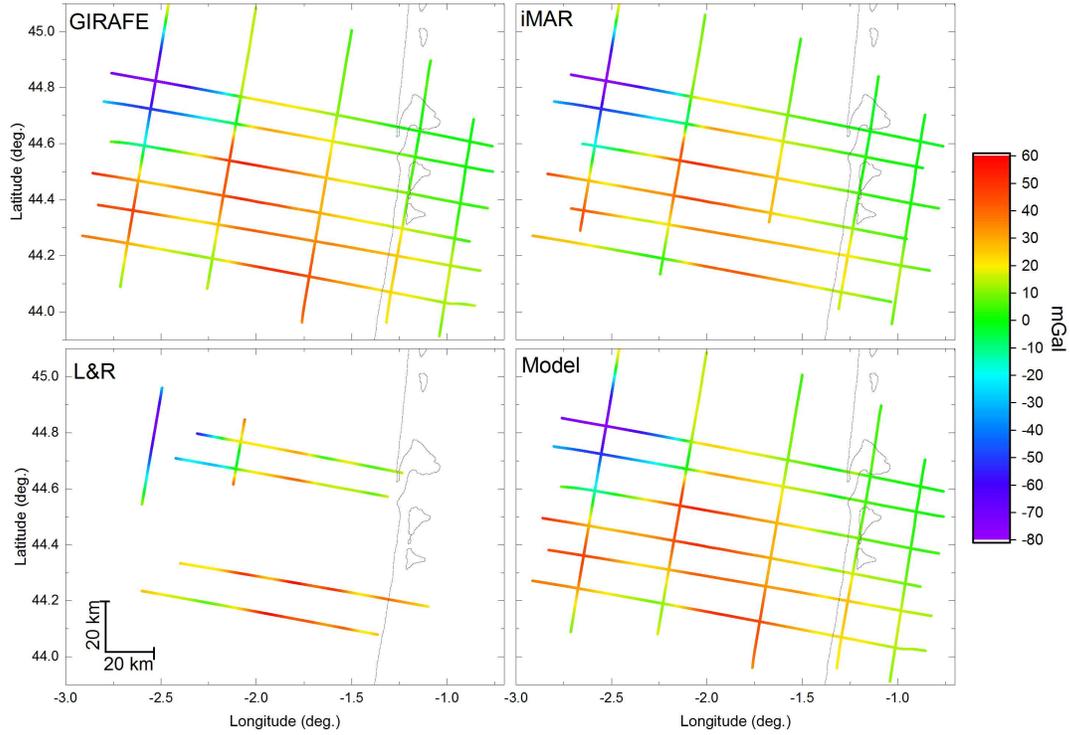}
\caption{Gravity disturbance over Biscay Bay measured by the three gravimeters and model at the flight altitude derived from satellite altimetry and land gravimetry data (see section \ref{section model}). The filter used for GIRAFE data processing leads to a FWHM spatial resolution of 7 km (see section \ref{GIRAFE processing}). The missing L\&R data are due to excessive turbulence.}
\label{biscay}
\end{figure}

\begin{figure}[h!]
\vspace{-2cm}
 \noindent\includegraphics[width=20 cm,natwidth=2936px,natheight=1095px]{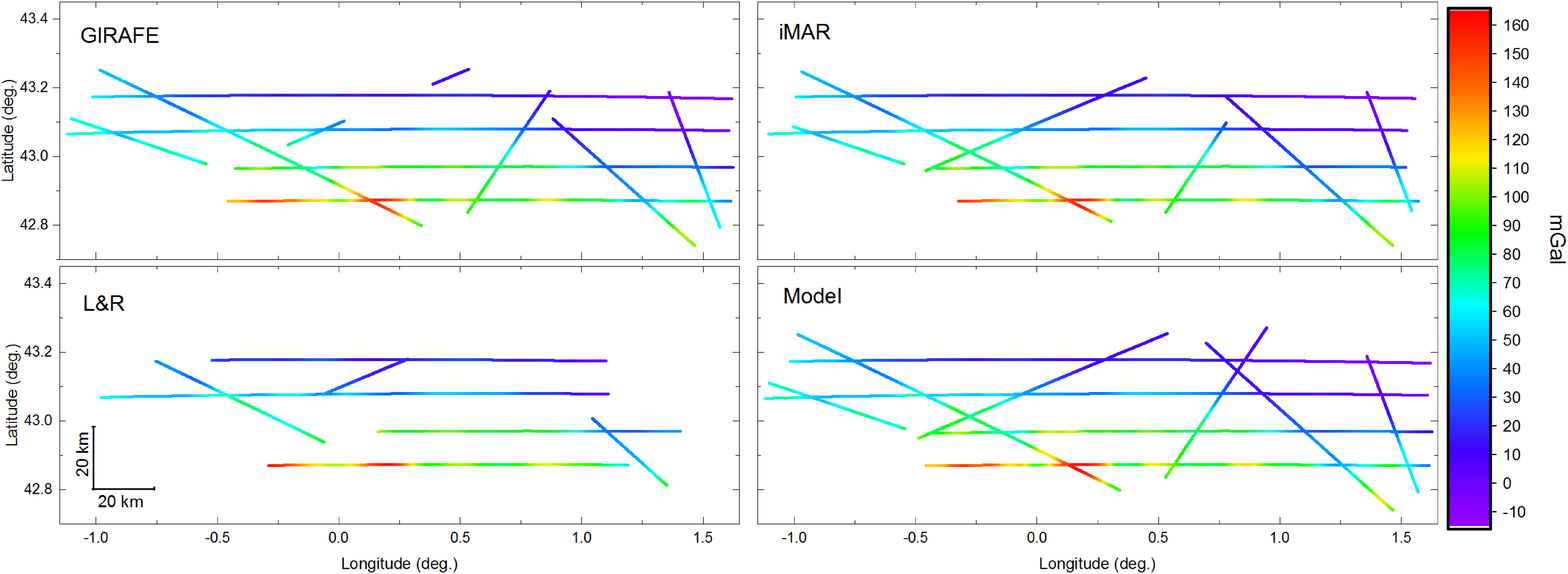}
\caption{ Gravity disturbance over Pyrenees measured by the three gravimeters and gravity models at the flight altitude (see section \ref{section model}). The filter used for GIRAFE data processing leads to a FWHM spatial resolution of  4.5 km (see section \ref{GIRAFE processing}).  The model values are from upward continued land gravity data in France and Spain.}
\label{pyrenees}
\end{figure}

\begin{figure}[h!]
\vspace{-2cm}
 \noindent\includegraphics[width=18 cm,natwidth=3008px,natheight=2175px]{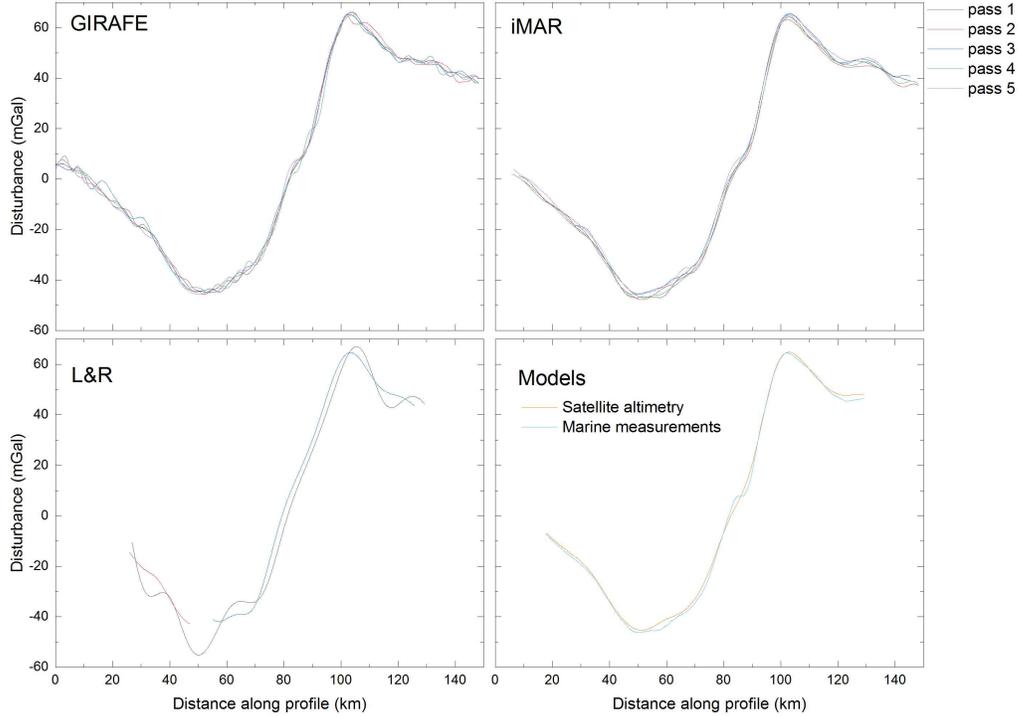}
\caption{Gravity disturbance along the reference profile measured by the three gravimeters and gravity models at the flight altitude (see section \ref{section model}).The filter used for GIRAFE data processing leads to a FWHM spatial resolution of 4.5 km (see section \ref{GIRAFE processing}).}
\label{ref_profile}
\end{figure}

All data have been processed to compute the gravity disturbances. Detail of the data processing used to derive gravity disturbance from gravimeters measurements and GNSS data is described in \ref{DataProc}. The final maps of disturbances estimated from the three gravimeters along the flight lines at flight altitude are given on Figures \ref{biscay} and \ref{pyrenees} for Bay of Biscay and Pyrenees areas respectively on Figure \ref{ref_profile} for the reference profile. Turns or lines with bad data acquisition for the different gravity meters used during surveys were removed.  Note that laser frequency locking problems occurred on GIRAFE gravimeter on May 24 during Pyrennes flight, prevented us to acquire data along certain lines. The gravity disturbance derived from the marine, land and altimetry gravity data projected on the flight lines are also given for reference. At a glance, we can see that GIRAFE and iMAR mostly show comparable acquisition lines while the L\&R meter led to less exploitable data acquisition. Gravity anomaly patterns and amplitudes revealed by all meters are close to the expected ones derived from marine, land and altimetry data.  The gravity disturbances obtained from the three gravity sensors along the flight lines are analyzed hereafter in order to estimate the measurement errors and the stability.

\FloatBarrier

\section{Estimation of the precision of the gravimeters from repeated measurements}

\subsection{Measurement errors}

\begin{figure}[h!]
\begin{center}
\vspace{-0cm}
 \noindent\includegraphics[width=19 cm,natwidth=2611px,natheight=2371px]{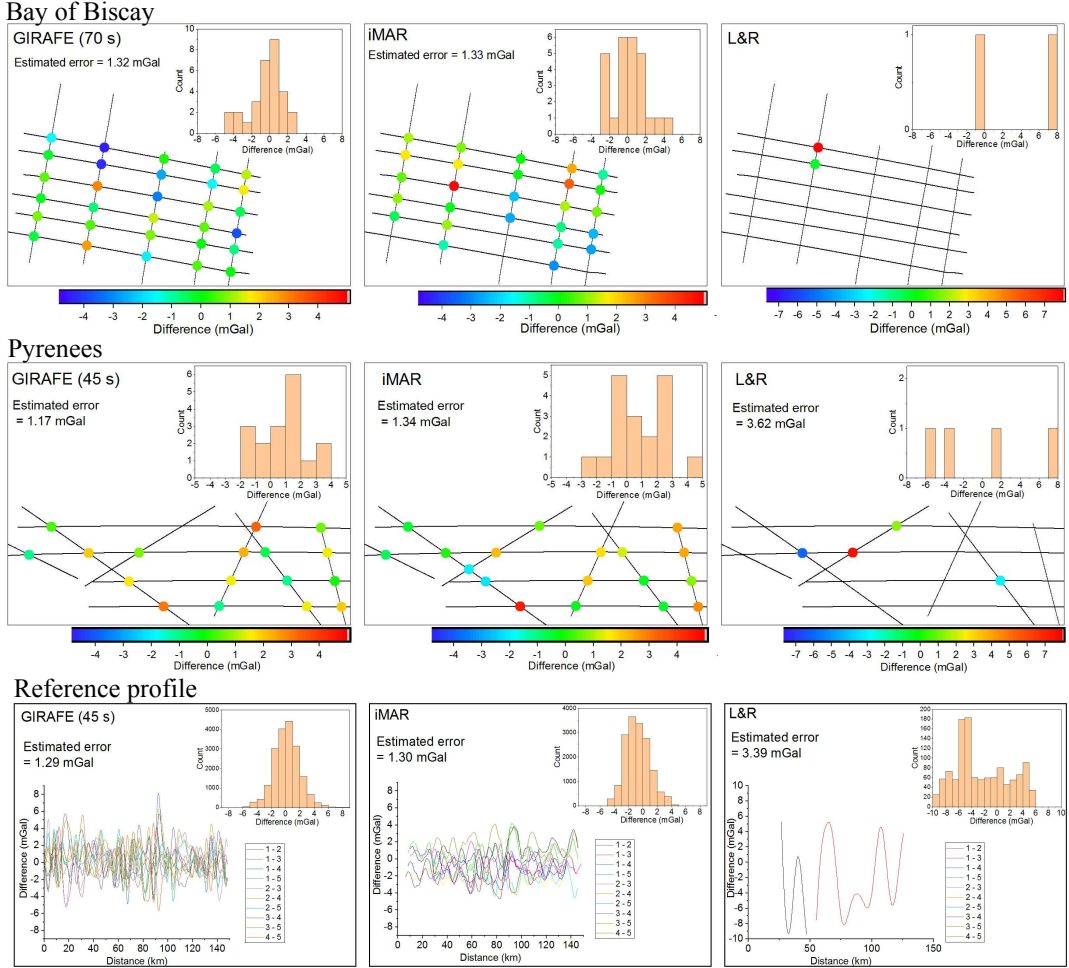}
\end{center}
\caption{Crossing points analysis and repeated profile analysis. The first and second row show the crossing point differences for Bay of Biscay and Pyrenees survey. The last row shows the differences between the 5 repeated lines over the reference profile. The curves labelled i-j represent the difference between the gravity measured at i-th pass and j-th pass. On each graph, the upper-right insert is the histogram of the differences showed on the graph. The FWHM pulse response $\Delta\tau$ of the filter used in GIRAFE data processing is 70 s for Bay of Biscay, 45 s for Pyrennes and reference profile.}
\label{crossing points}
\end{figure}

The error of the gravity measurements can be estimated by analyzing the differences between gravity measurements performed at the same location. For the reference profile, repeated measurements over the same line are compared and for the Bay of Biscay and Pyrenees, crossing point differences are analyzed (see Figure \ref{crossing points}). Note that the cross-over analysis has not been done for L\&R gravimeter in Biscay Bay because there are only two crossing points.

Assuming uncorrelated errors between measurements, the estimated error can be computed using the following expression :
\begin{equation}
\epsilon_{\delta g} = \frac{1}{\sqrt{2}}\sqrt{\frac{1}{N} \sum_{\substack{n,i,j \\ i>j}} \left(\delta g_i (r_n)-\delta g_j(r_n)\right)^2 }
\end{equation}
where $\delta g_i(r_n)$ is the gravity disturbance measured at the position $r_n$ at the i th pass and $N$ is the number of elements in the sum. 

\begin{figure}[h!]
\vspace{-2cm}
 \noindent\includegraphics[width=15 cm,natwidth=3339px,natheight=2215px]{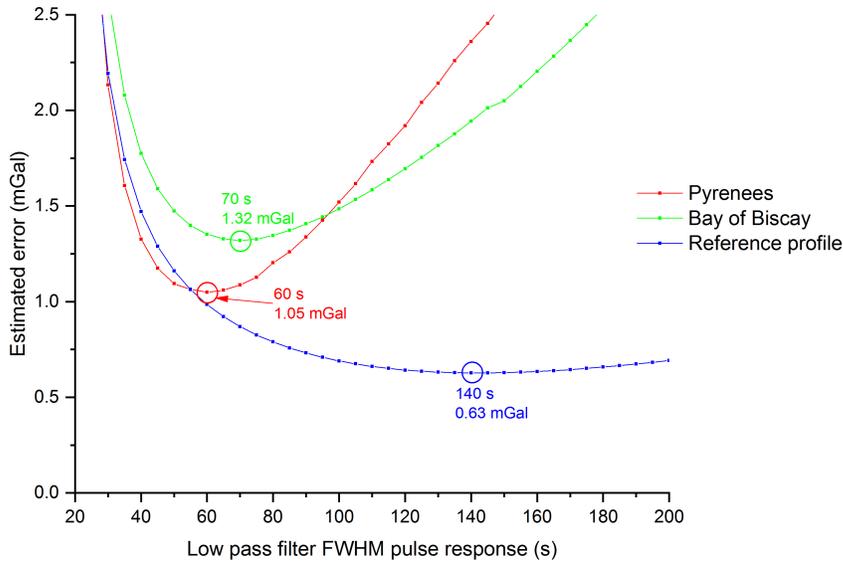}
\caption{Estimated error on GIRAFE gravity measurements versus the FWHM pulse response of the low pass filter $\Delta\tau$ used in the data processing.}
\label{resolution}
\end{figure}

\FloatBarrier
The influence of the cutoff frequency of the used Gaussian low pass filter in the GIRAFE gravimeter data processing has been studied (see \ref{GIRAFE processing}). In the following, the filter will be characterized by its FWHM pulse response $\Delta\tau$. The spatial resolution of the measurements i.e. the FWHM response to an infinitely narrow gravity anomaly peak (Dirac delta function) is then equal to $v\times\Delta\tau$ with $v$ the plane velocity. On Figure \ref{resolution}, we plot the estimated errors for the three surveys versus the FWHM pulse response $\Delta\tau$. For the three surveys, there is a value of $\Delta\tau$ which minimizes the error.  However, the optimal $\Delta\tau$ is different for each survey. For small values of $\Delta\tau$, the estimated error is smaller for Pyrenees and reference profile than for Biscay bay. This could be explained by the fact that Biscay flights were the most turbulent leading to a larger measurement noise. For large values of $\Delta\tau$ the estimated error is smaller for the reference profile than Bay of Biscay and Pyrenees. This can be explained first by the fact that the measurement line on the reference profile is longer leading to less border effects, but also due to the differences in turbulence. Second, the error is estimated by comparing measurements acquired on the same line for the reference profile and on different lines for Biscay bay and Pyrenees. As the low pass filter make a spatial filtering only on a 1D line, spatially unresolved gravity anomaly can lead to different gravity measurements if differently oriented measurement lines are used. This directional effect is especially important in the Pyrenees flights, where the local gravity variations are much higher, and more anisotropic, than in the marine flights.

On Table \ref{Errors}, the estimated errors for the three gravimeters and for each area is summarized. In order to have a meaningful comparison, the low pass filter used for GIRAFE data processing has been selected to minimize the rms differences with iMAR measurements (see section \ref{section gravi comp}). If we assume that both gravimeters have the same noise level, this choice leads to the same spatial resolution.

GIRAFE and iMAR have approximately the same measurement errors ranging from 1.2 to 1.3 mGal. On the other side, L\&R gravimeter has larger errors ranging from 3.4 to 3.6 mGal. Indeed, the L\&R platform is quite sensitive to turbulence and dynamic flight conditions.

\begin{table}[h!]
 \caption{Gravity measurement errors estimated from repeated measurements (mGal). }
 \centering
 \begin{tabular}{l l l l l }
														&	Bay of Biscay						& Reference profile						& Pyrenees		\\
 \hline
GIRAFE quantum gravimeter		& 1.32 										& 		1.29 										& 1.17 			\\
														&($\Delta\tau$=70s)				& ($\Delta\tau$=45s)					& ($\Delta\tau$=45s)\\
\hline
iMAR strapdown gravimeter	  & 1.33										& 		1.30										& 1.34				\\

\hline	
L\&R platform gravimeter		& undetermined					  & 		3.39					 					& 3.62					\\
									
 \end{tabular}
\label{Errors}
 \end{table}

\FloatBarrier

\subsection{Stability of gravity measurements}

\begin{figure}[h!]
\vspace{-2cm}
 \noindent\includegraphics[width=14 cm,natwidth=3431px,natheight=2276px]{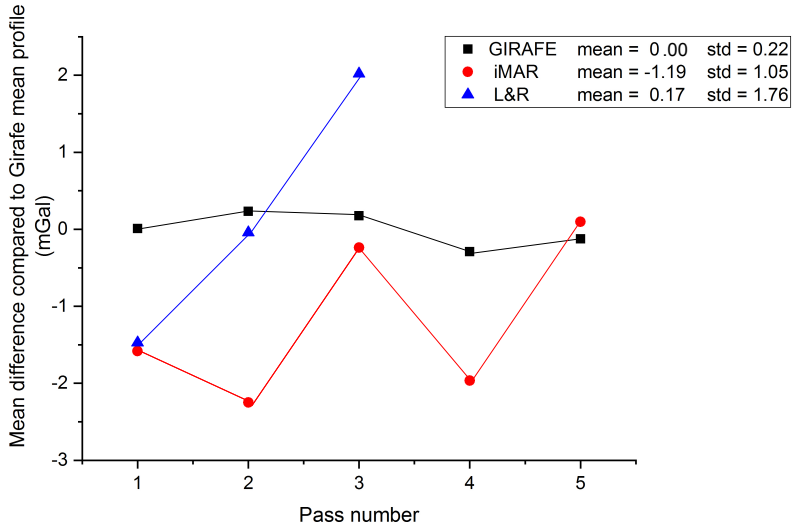}
\caption{Stability of gravity measurements over the reference profile.}
\label{stability}
\end{figure}

The stability of the gravity measurements during the flight is estimated by comparing the mean values of the gravity disturbance measured at each pass over the reference profile (see Figure \ref{stability}). The standard deviation on the gravity mean values is clearly smaller for GIRAFE gravimeter (0.22 mGal) than for iMAR (1.05 mGal) and L\&R (1.76 mGal). This confirms the advantage of GIRAFE absolute gravimeter over relative gravimeters even if these relative gravimeters have been calibrated and the drift is compensated for.

\FloatBarrier 

\section{Comparison of gravity disturbance estimated by the three gravimeters}\label{section gravi comp}

\begin{figure}[h!]
\vspace{-5cm}
  \noindent\includegraphics[width=18 cm,natwidth=3040px,natheight=2164px]{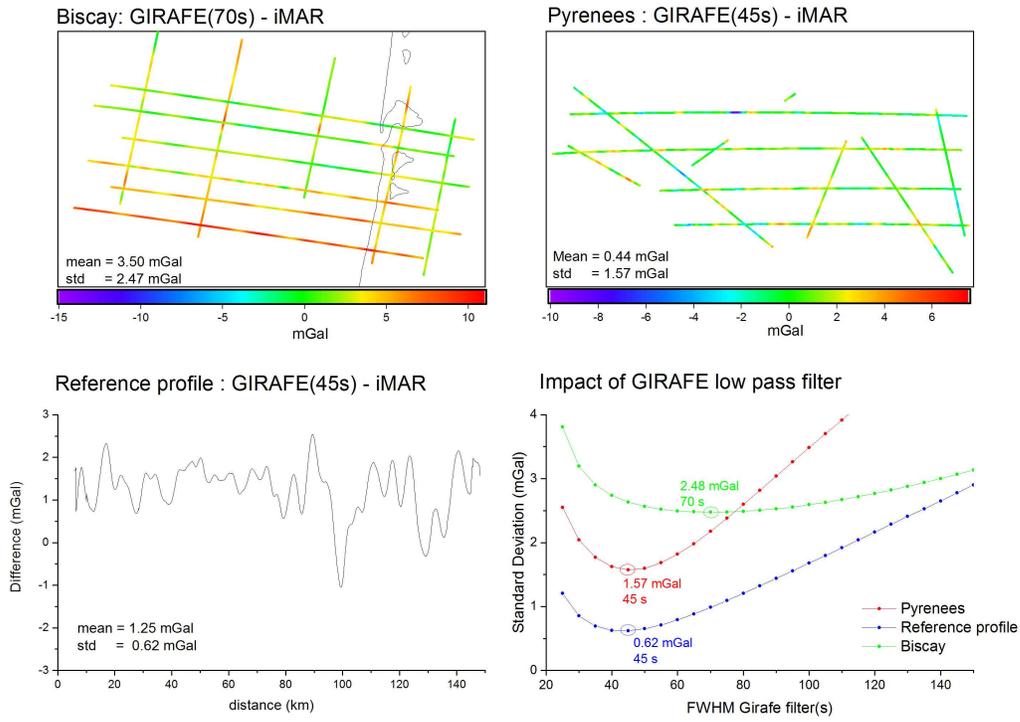}
\caption{Comparison of gravity disturbance estimated by GIRAFE and iMAR gravimeters. The two upper graphs represent the difference between GIRAFE and iMAR gravity measurements for Bay of Biscay and Pyrennees. The lower-left graph is the difference between the averaged gravity measurements over the reference profile for GIRAFE and iMAR. The lower-right graph shows the standard deviation of the difference between GIRAFE and iMAR measurements versus the filter FWHM pulse response $\Delta\tau$ used for GIRAFE data processing.}
\label{comp girafe imar}
\end{figure}

\begin{figure}[h!]
\vspace{0cm}
  \noindent\includegraphics[width=18 cm,natwidth=3075px,natheight=2099px]{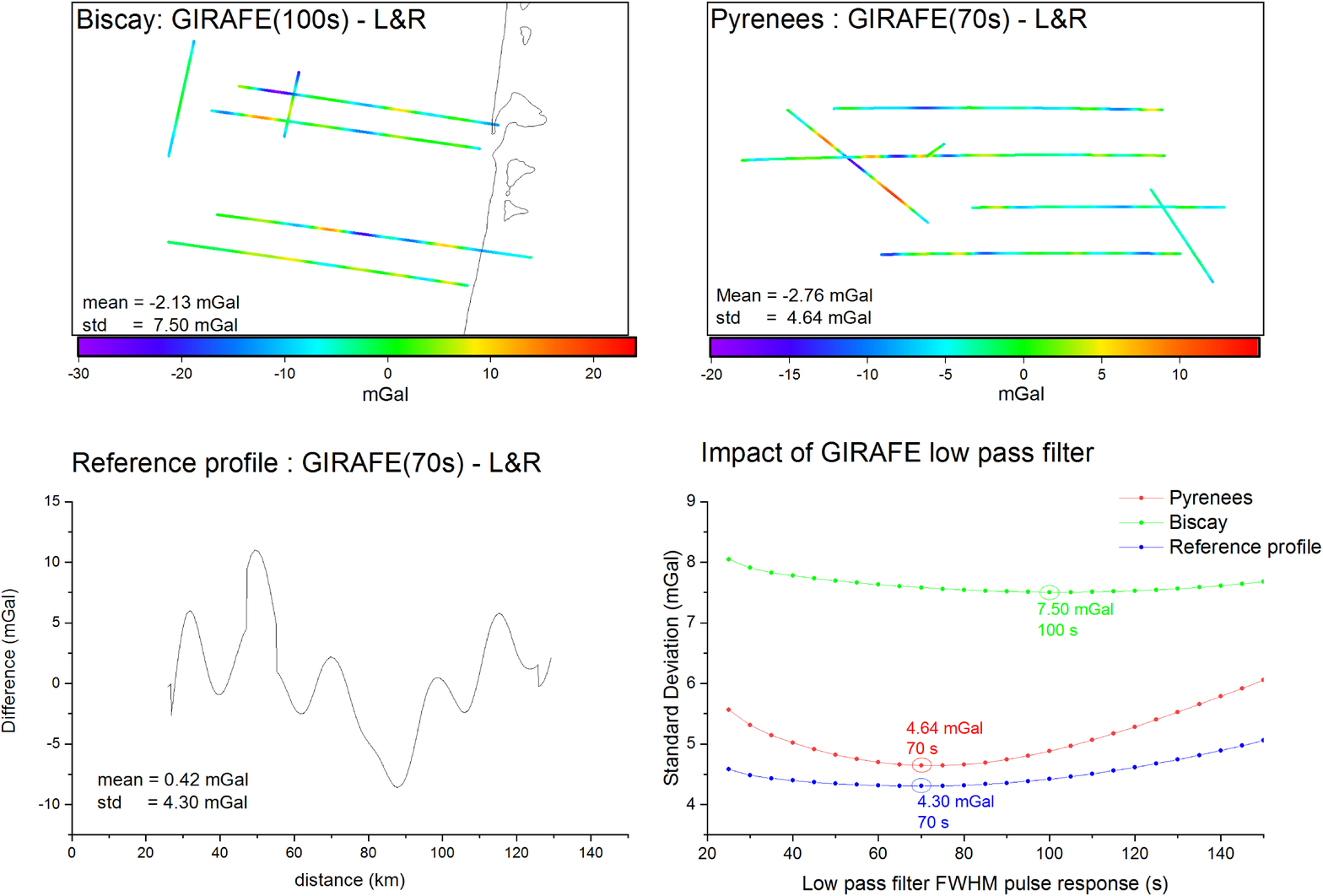}
\caption{Comparison of gravity disturbance estimated by GIRAFE and L\&R gravimeters. The two upper graphs represent the difference between GIRAFE and L\&R gravity measurements for Bay of Biscay and Pyrennees. The lower-left graph is the difference between the averaged gravity measurements over the reference profile for GIRAFE and L\&R. The lower-right graph shows the standard deviation of the difference between GIRAFE and L\&R measurements versus the filter FWHM pulse response $\Delta\tau$ used for GIRAFE data processing.}
\label{comp girafe LCR}
\end{figure}

The gravity disturbance estimated from the measurements of the three gravimeters has been compared in Figure \ref{comp girafe imar} and \ref{comp girafe LCR}. For the reference profile, the gravity disturbance is averaged over 5  passes for GIRAFE and iMAR and over 2 passes for L\&R. 

We study here also the influence of the low pass filter used for GIRAFE data processing (see appendix \ref{GIRAFE processing}). This low pass filter acts on the measurement noise and on the spatial resolution. We observe that a filter FWHM pulse response $\Delta\tau$ minimizes the standard deviation of the difference between gravimeter measurements. For iMAR, the minimum is 45 s for Pyrenees and reference profile and 70 s for Bay of Biscay. For L\&R, the minimum is 70 s for Pyrennes and reference profile and 100 s for Bay of Biscay. Those minimums result in a compromise between spatial resolution matching between gravity measurements and GIRAFE measurement noise minimizing. Neglecting GIRAFE measurement noise, the minimum corresponds to the point where the measurements of both gravimeters have the same spatial resolution. 

The standard deviations on the differences for those minimums are summarized in Table \ref{Comp_std_gravi}. They are compared to the estimation obtained by summing quadratically  each gravimeter errors calculated from repeated measurements at the same location (see Table \ref{Errors}). For the reference profile, the error on the averaged profile is equal to the error divided by the square-root of the number of pass. For GIRAFE/iMAR comparison,  the same values are obtained within 30\%  validating the estimated error for each gravimeters. For L\&R, the comparison of the two approaches was only possible for Pyrenees and reference profile where it was possible to estimate errors from repeated measurements and leads to less good agreement.

\begin{table}[h!]
 \caption{Standard deviation on the difference between gravimeter measurements (mGal). The values in parenthesis are the expected standard deviation obtained by summing quadratically each gravimeter errors.}
 \centering
 \begin{tabular}{l l l l l }
														&	Bay of Biscay			& Reference profile		& Pyrennees		\\
 \hline
GIRAFE -iMAR								& 2.47	(1.87)			& 		0.62 (0.82)			& 1.57		(1.78)	\\
											
\hline
GIRAFE - L\&R							  & 7.50							& 	  4.30	(2.47)			& 4.64		(3.80)		\\

\hline
iMAR - L\&R									& 7.72						  &     4.36	(2.47)			& 5.09		(3.86)		\\
\hline									
 \end{tabular}
\label{Comp_std_gravi}
 \end{table}

\FloatBarrier 
The differences in mean value between the gravity measurements are summarized in Table \ref{Comp_mean_gravi}. The difference between iMAR and GIRAFE are in the mGal level for reference profile and Pyrennees (1.25 mGal, 0.44 mGal). However, the difference is larger (3.5 mGal) for Bay of Biscay. The comparison with gravity models (see \ref{section model biscay}) suggests that the iMAR estimates are subject to an offset. Concerning L\&R gravimeter the differences are below 1 mGal for the reference profile and between 2 and 6 mGal for Pyrennes and Bay of Biscay.

\begin{table}[h!]
 \caption{Comparison between mean values of gravimeter measurements in mGal. }
 \centering
 \begin{tabular}{l l l l l }
														&	Bay of Biscay			& Reference profile		& Pyrennees		\\
 \hline
GIRAFE -iMAR								& 3.50							& 		1.25						& 0.44				\\
											
\hline
GIRAFE - L\&R							  & -2.13					   	& 	  0.42						& -2.76			\\

\hline
iMAR - L\&R									& -6.30						  &     -0.85					 	& -3.13					\\
\hline									
 \end{tabular}
\label{Comp_mean_gravi}
 \end{table}

\FloatBarrier 
\section{Comparison with upward continued-surface gravity and altimetry-derived gravity}\label{section model}

We use the upward continued-surface gravity as an independent data to validate the airborne gravity measured by the three gravimeters.

\subsection{Reference profile} \label{part comp_airborne_marine_ref_profile}

For the reference profile, the airborne measurements have been compared with Shom marine data acquired in 2018 with the quantum gravimeter GIRAFE, and also with gravity model from satellite altimetry V31 \cite{Sandwell2014}. Marine data and altimetry model have been upward continued to flight altitude using the spectral method of \cite{Blakely1996}. In addition, to bring them back to the airborne measurement height reference (WGS84 ellipsoid), the marine data were also corrected to gravity disturbances by the geoid height of the EGM08 model at order 2190 \cite{Pavlis2012}.

\begin{figure}[h!]
	\vspace{-3cm} 
 \noindent\includegraphics[width=19 cm,natwidth=3212px,natheight=2199px]{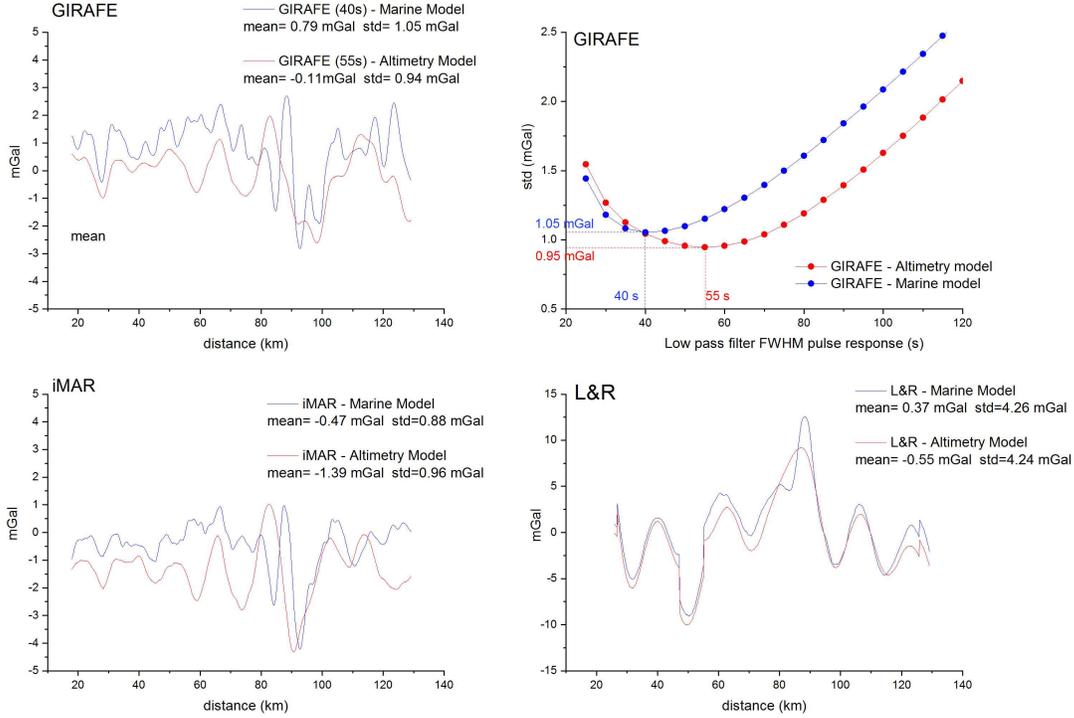}
\caption{Comparison between airborne measurements and two reference models (marine gravimetry and satellite altimetry). The upper-left graph is the differences between GIRAFE measurements and the two models. The upper-right graph is the standard deviation of the differences between GIRAFE measurements and models versus the FWHM pulse response $\Delta\tau$ of the filter used for GIRAFE data processing. The lower-left graph is the differences between iMAR measurements and the two models. The lower-right graph is the differences between L\&R measurements and the two models.}
\label{fig_comp marine data}
\end{figure}

The results of the comparison are shown in Figure \ref{fig_comp marine data}.  The airborne gravity disturbance are obtained by averaging over the 5 passes for iMAR and GIRAFE gravimeters and over 2 passes for L\&R.

Again, the influence of the FWHM pulse response of the filter $\Delta\tau$ used for GIRAFE data processing was studied. For marine data, the standard deviation on the differences is minimum for $\Delta\tau$=40 s and is equal to 1.05 mGal. For the altimetry model, the standard deviation on the differences is minimum for $\Delta\tau$=55 s and is equal to 0.95 mGal.  The difference of the minimum values could be explained by the higher resolution of the marine gravity data.

GIRAFE and iMAR measurements are in good agreements with gravity models. The differences between GIRAFE have a mean and a standard deviation below or approximately equal to 1 mGal. We notice that the standard deviation between airborne measurements and models (0.9 - 1 mGal) is bigger than the standard deviation between iMAR and GIRAFE (0.6 mGal). This could be explained by errors on gravity models or common errors on iMAR and GIRAFE measurements like for example GNSS errors.

For L\&R, we obtain a good agreement for the mean value. The standard deviation on the difference is however bigger (4 mGal) confirming larger measurement errors for L\&R compared to iMAR and GIRAFE.

\FloatBarrier 

\subsection{Bay of Biscay and Pyrenees} \label{section model biscay}
	
The upward continuation estimation of the gravity disturbances at the flight altitude was performed in point-to-point mode with the well-known remove-compute-restore technique. First, the gravity disturbances from the EGM2008 up to degree/order 2190 \cite{Pavlis2012} were used to remove/restore the long wavelength components in the gravity data. We also used the Residual Terrain Model (RTM) effects \cite{Forsberg1984} computed from Digital Terrain Model (DTM) to remove/restore the short wavelengths (beyond degree/order 2190). On the Pyrenees, the 90 m resolution SRTM3arc v4.1 \cite{Farr2007} was used as the detailed DTM to compute the RTM effects. The 15” resolution Digital Bathymetry Model (DBM) SRTM15arc plus \cite{Tozer2019} was merged with SRTM3arc V4.1 to compute the RTM effects on the Bay of Biscay. Then, the residual gravity disturbances have been upward continued to the flight altitude using the Least-Squares Collocation (LSC) \cite{Forsberg1987}.

\begin{figure}[h!]
	\vspace{0cm} 
 \noindent\includegraphics[width=19 cm,natwidth=1258px,natheight=859px]{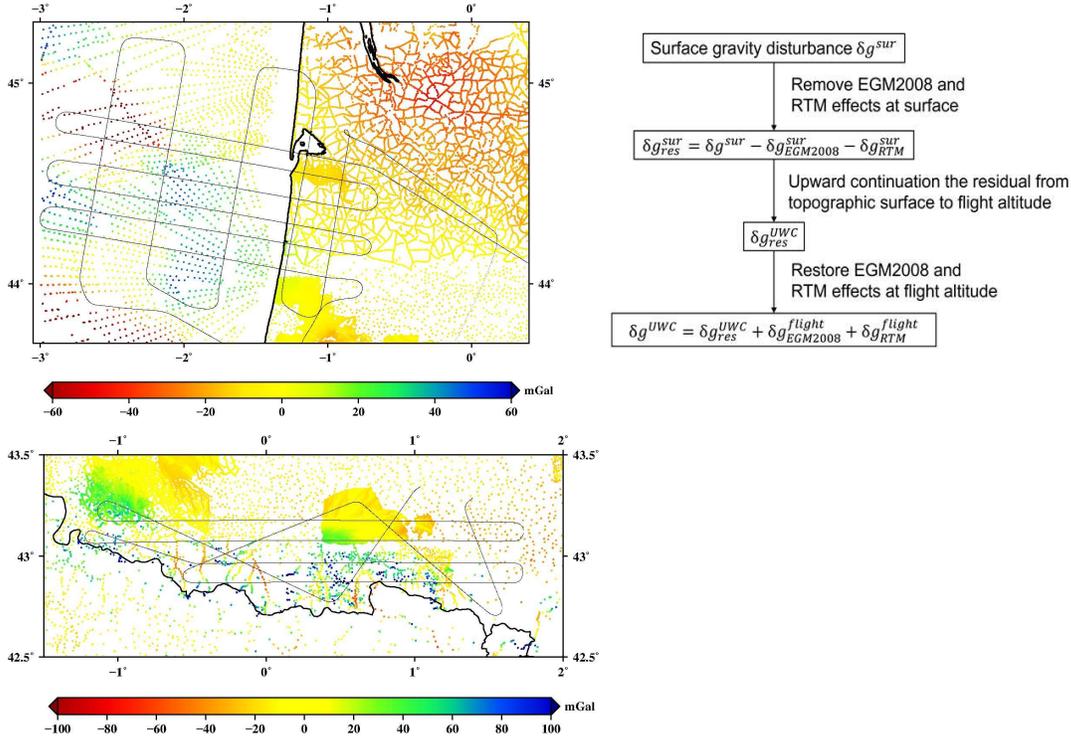}
\caption{Upper Left : Terrestrial and marine gravity data in Bay of Biscay. Lower left: Terrestrial data in Pyrenees. Upper right : Diagram of sequential steps (top to bottom) in the upward continuation of gravity disturbance.}
\label{fig_method model}
\end{figure}

\begin{figure}[h!]
	\vspace{-3cm} 
 \noindent\includegraphics[width=16 cm,natwidth=2403px,natheight=2516px]{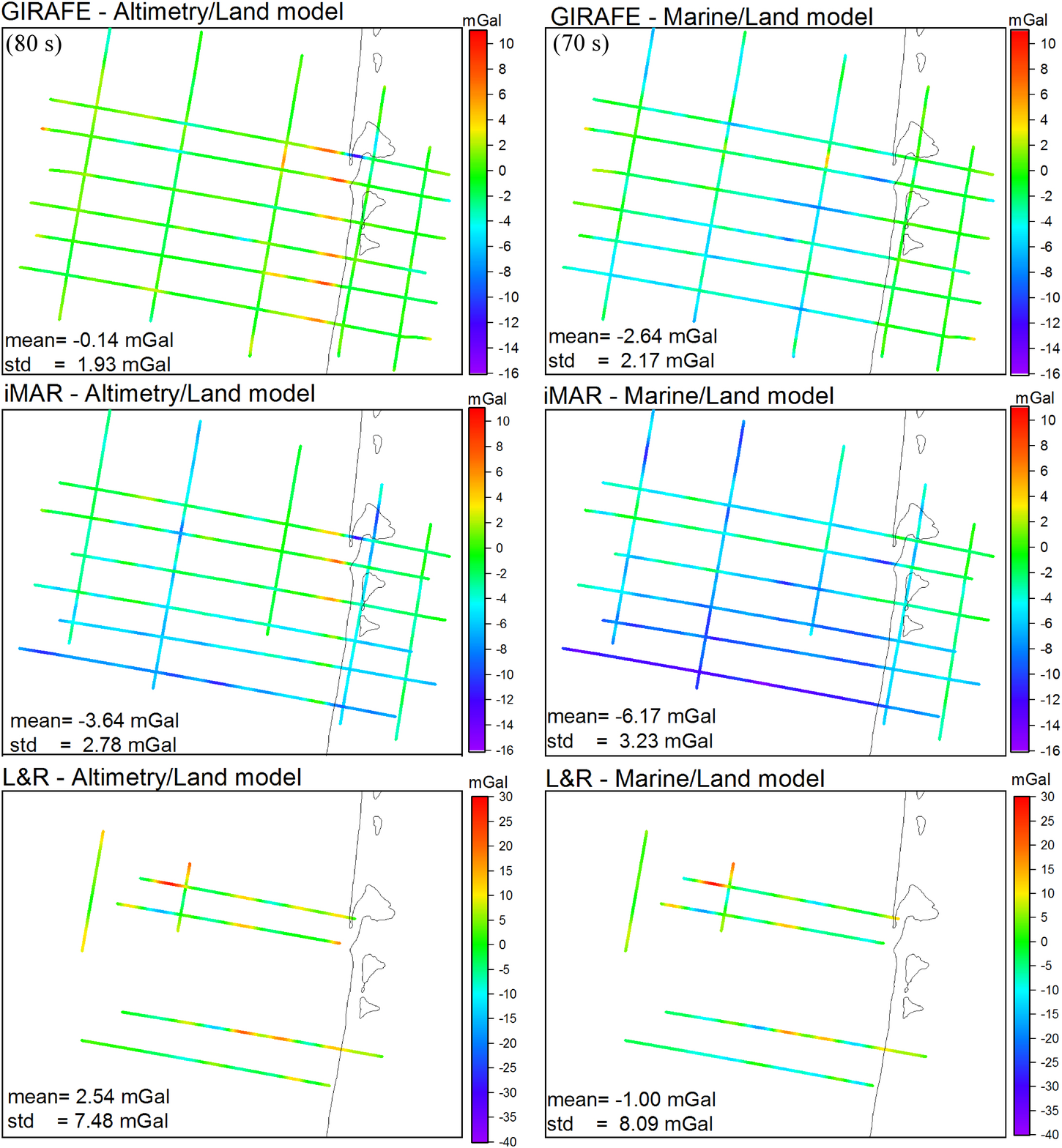}

\vspace{-1.5cm}

\noindent\includegraphics[width=10 cm,natwidth=603px,natheight=444px]{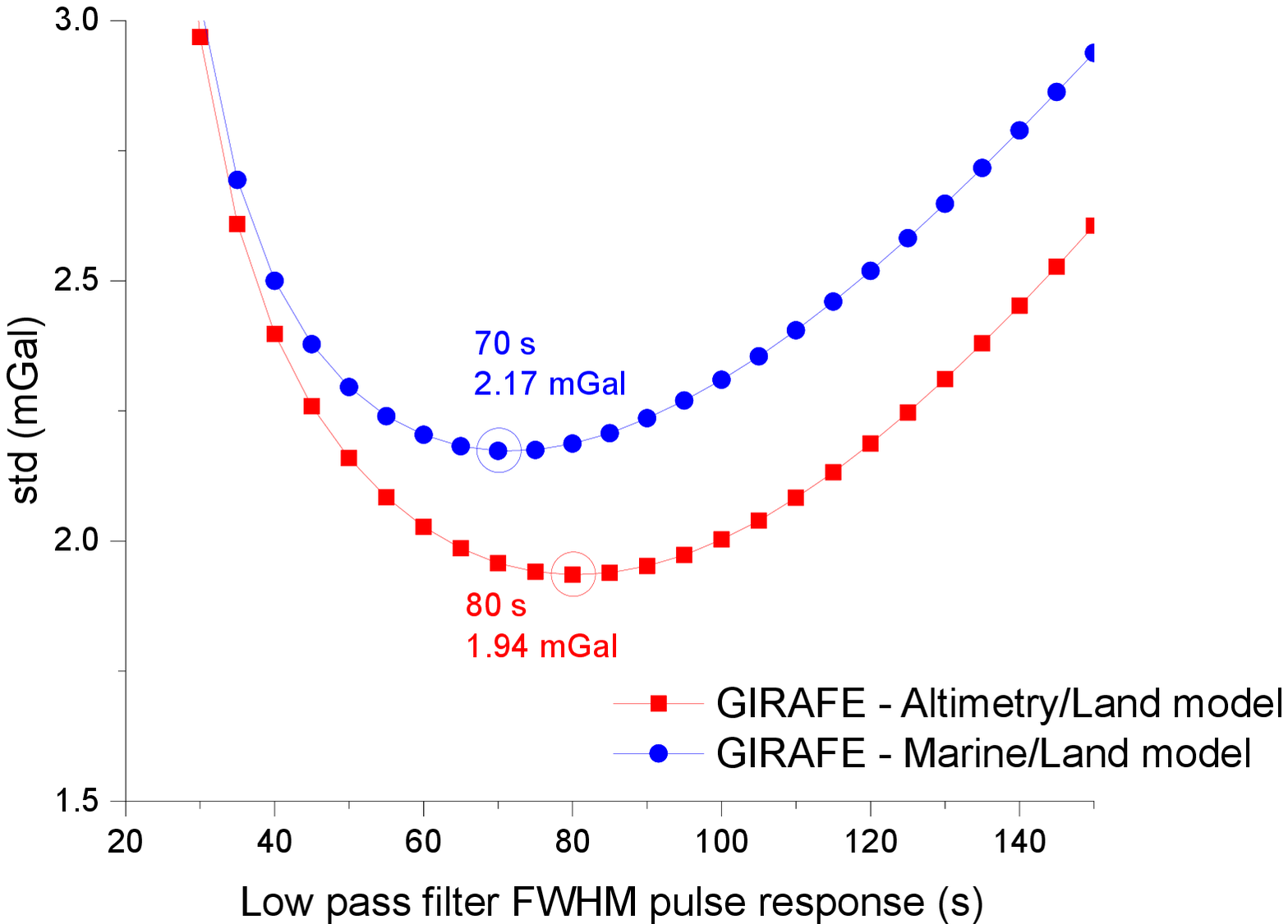}

\caption{Comparison between gravity disturbance over Bay of Biscay derived from airborne measurements and from gravity models. The six upper graphs represent the difference between airborne measurements (GIRAFE, iMAR or L\&R) and models (Satellite altimetry/Land or Marine/Land). The lowest graph represents the standard deviation of the differences between GIRAFE measurements and models versus the FWHM pulse response of the filter $\Delta\tau$ used for GIRAFE data processing.} 
\label{fig_comp model biscay}
\end{figure}

\begin{figure}[h!]
	\vspace{-3cm} 
 \noindent\includegraphics[width=18 cm,natwidth=2759px,natheight=1695px]{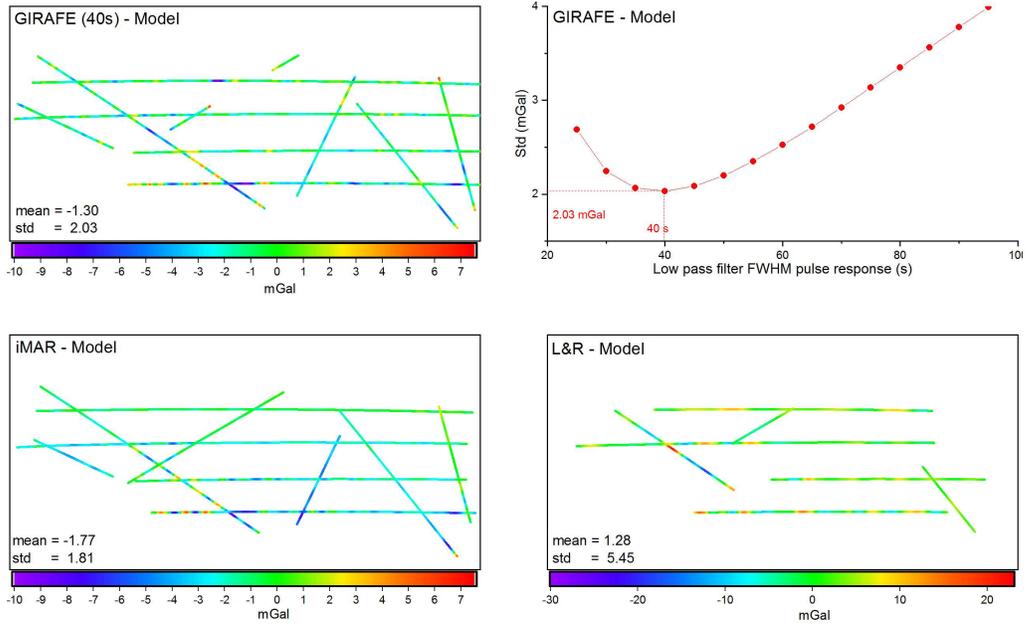}
\caption{Comparison between gravity disturbances over Pyrenees derived from airborne measurements and from land measurements upward continued. The upper-left graph is the difference between GIRAFE measurements ($\Delta\tau$=40s) and the model. The upper-right graph is the standard deviation of the differences between GIRAFE measurements and model versus the FWHM pulse response of the filter $\Delta\tau$ used for GIRAFE data processing. The two lower graphs are the difference between iMAR and L\&R measurements and the model.}
\label{fig_comp model pyrenees}
\end{figure}

In Figure \ref{fig_method model}, $\delta g^{sur}$ and $\delta g^{sur}_{res}$ denote surface gravity disturbances (land, marine measurements or altimetry model) and their residual values, respectively.  $\delta g^{sur}_{EGM2008}$ and $\delta g^{flight}_{EGM2008}$  are the long-wavelength gravity disturbances at the topographic/sea surface and at the flight altitude, respectively. $\delta g^{sur}_{RTM}$ and $\delta g^{flight}_{RTM}$ represent the topographic gravity effects at the topographic/sea surface and at the flight altitude, respectively. Therefore, the long-wavelength components and the topographic gravity are computed at different altitude in remove and restore procedures. $\delta g^{UWC}$ and $\delta g^{UWC}_{res}$ are upward continued-gravity disturbances and their residual values, respectively. 
The terrestrial gravity data were provided by the International Gravimetric Bureau (BGI) and GETECH while the marine gravity data provided by BGI. Fortunately, two study regions have a relatively dense gravity coverage, as shown in Figure \ref{fig_method model}. The differences between upward continued and the airborne gravity disturbances measured by three gravimeters are shown in Figure \ref{fig_comp model biscay} and \ref{fig_comp model pyrenees}.

For the Bay of Biscay, the standard deviation (STD) and mean value of these differences are clearly smaller for GIRAFE than for iMAR and L\&R. We obtained a standard deviation on the differences equal to 2.17 mGal and a mean bias of -2.64 mGal for GIRAFE. A large bias is visible over the sea part relative to the land, which leads the quite bias in the validation result on the Bay of Biscay. To make this clear, we also use the gravity field derived from the latest altimetric satellite model, namely V31, to validate the airborne gravity. Similar to marine gravity, the gravity disturbance derived from V31 model \cite{Sandwell2014} is also upward continued to the flight altitude. The bias is no longer visible on the marine part, the mean bias being -0.4 mGal on the Bay of Biscay for GIRAFE. The bias is  larger for iMAR and L\&R (-3.64 and 2.54 mGal respectively). This confirms once again the advantage of GIRAFE absolute gravimeters over relative gravimeters. 

The differences between the upward continued using altimetric gravity and the airborne gravity also reveal that the altimetric gravity is of poor quality in the coastal areas (about 10 km from the coastal line). This explains why the airborne gravimetry is used preferentially to close the gap between gravity data on land and marine altimetric gravity fields.
For the Pyrenees, the STD is around 2 mGal for the GIRAFE and iMAR measurements. This indicates that these two measurements are in good agreements with the upward continued-ground gravity disturbances. The L\&R measurements have large differences from the upward continued-ground gravity (STD is 5.45 mGal). The smallest mean bias obtained from GIRAFE (-1.30 mGal), however, this value is quite large. The reason for the large mean bias are in part the reference height inconsistencies, but also inconsistencies in the atmospheric corrections of gravity data, and errors in the upward continuation process. The ground gravity measurements are thus referring to the national height systems, while the airborne gravity is calculated using the height referring to a global height system. Using more than 10.000 GPS/levelling points in France, we have determined that the vertical datum offset of the national height system of France is about 0.87 m (equivalent to 0.3 mGal) with respect to the international (W0) height system \cite{Sanchez2016}; atmosphere effects are up to 0.8 mgal, and additional errors from linear corrections terms in BGI/GTECH may also contribute to the bias. It is therefore quite likely that the GIRAFE bias might be close to zero also over the mountains.

\FloatBarrier 

\section{Conclusion and perspectives}

\begin{figure}[h!]
	\vspace{-1cm} 
 \noindent\includegraphics[width=22 cm,natwidth=2763px,natheight=788px]{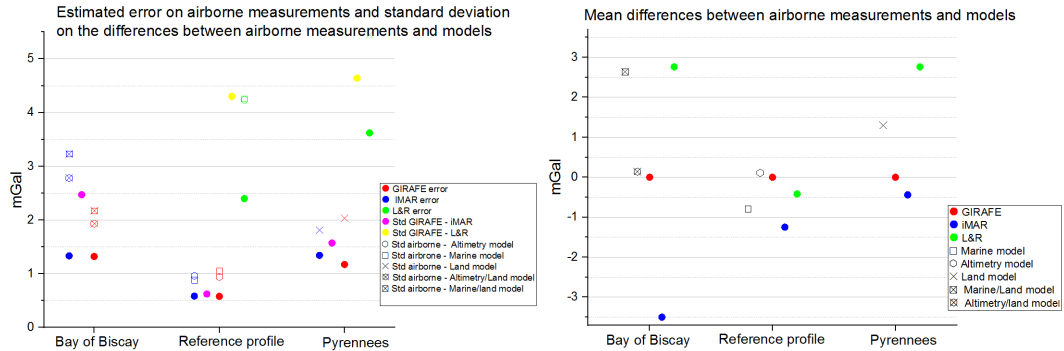}
\caption{Summary of the intercomparison between GIRAFE, iMAR and L\&R airborne gravity measurements and gravity models. Left: Estimated error on airborne measurements and standard deviation on the differences between airborne measurements and models. Right : Mean differences between airborne measurements and models. The reference value is GIRAFE measurements.}
\label{fig_summary}
\end{figure}

In conclusion, we demonstrated absolute airborne gravity measurements with a quantum sensor. From repeated measurements, we estimated measurements errors ranging from 0.6 to 1.3 mGal. Precision measurements have been improved by a factor 3 compared to the first airborne campaign with the quantum gravimeter. 
Gravity measurements from two relative gravimeters were acquired simultaneously. The comparison indicated a similar precision between the GIRAFE and iMAR strapdown instruments, whereas the L\&R platform system performed poorly in the present flight conditions. We also observed that the long term stability of GIRAFE gravimeter is 5 times better than the iMAR gravimeter. The measurements of the three gravimeters have been compared to gravity models coming from satellite altimetry, marine and ground measurements. A summary of the comparisons is shown on Figure \ref{fig_summary}. Good agreements are obtained between GIRAFE, iMAR and gravity models. 

This survey confirms the potential of quantum sensor for airborne gravimetry. Better precision could be achieved in the future with quantum technology. The sensitivity and the accuracy of the quantum sensor could still be improved. For example, we could use longer interrogation time, use a more sensitive auxiliary classical accelerometer or use multi-species atom interferometer \cite{Bonnin2018}. On the other hand, data processing has also to be improved. For example a Kalman filter approach could be used to process quantum gravity measurements. Moreover GNSS precision is expected to be improved in the future with the increasing of available constellations. With these improvements, we could hope to reach the precision to access to time variable gravity signals with application in earthquake, volcano, glacier or ground water changes.

The second point of improvement is the miniaturization of the sensors in order to have access to smaller carrier like drones. This could be achieved by miniaturizing the control cabinet and by designing, as for the iMAR gravimeter, a strapdown gravimeter which does not need gyrostabilised platform. This could be possible by gyrostabilizing the mirror retroreflecting the laser during the atom interrogation time \cite{Lan2012}. Three axes accelerometers combining one vertical and two horizontal atom accelerometers \cite{Perrin2019zero, Bernard2022} could be also of interest for strapdown gravimetry. This configuration could be also very promising for vectorial gravimetry where the horizontal components of the gravity field are also estimated.

\appendix

\section{Data processing and Gravity Estimation}\label{DataProc}

\subsection{Kinematic acceleration and E\"otv\"os effect}

Both classical and quantum gravimeters measure specific force, $f$, which is a combination of gravity acceleration, kinematic acceleration, and coupling to Earth rotation (E\"otv\"os effect). This may be expressed as

\begin{equation}\label{eq: sensor equation}
f = g + \ddot{h} - a_{\text{E\"ot}} \; ,
\end{equation}
where $g$ is the gravity acceleration ($g$ is positive when downward), $\ddot{h}$ is the time second derivative of $h$ the ellipsoidal height ($h$ is positive when upward) and represents the vertical kinematic acceleration of the aircraft. The centrifugal component due to Earth rotation has been absorbed by the gravity term, $g$. The E\"otv\"os term $a_{\text{E\"ot}}$ can be expressed as \cite{Heiskanen1967}:

\begin{equation}\label{eq: eotvos}
a_{\text{E\"ot}} = 2 \, \omega_E \, \cos\varphi \, v_{E} + \left[ \frac{v_{E}^2}{N(\varphi)+h}+\frac{v_{N}^2}{M(\varphi)+h} \right] \; ,
\end{equation}
with
\begin{equation}\nonumber
\begin{array} {lll}
\omega_E=7.292115\cdot 10^{-5}\,\text{s}^{-1}&:&\textnormal{\small Earth's rotation rate (inertial frame) }\\
\varphi &:&\textnormal{Geodetic latitude}\\
v_{E}&:&\textnormal{East velocity}\\
v_{N}&:&\textnormal{North velocity}\\
h&:&\textnormal{Ellipsoidal height}\\
M(\varphi)=\frac{a^2\cdot b^2}{\left(a^2\cos(\varphi)^2+b^2\sin(\varphi)^2\right)^{3/2}}&:& \textnormal{Earth's radius of curvature in the (north-south) meridian}\\
N(\varphi)=\frac{a^2}{\left(a^2\cos(\varphi)^2+b^2\sin(\varphi)^2\right)^{1/2}}&:& \textnormal{Earth's radius of curvature in the prime vertical}\\
a=6378137.0\,\meter&:&\textnormal{Earth's equatorial radius (WGS84)}\\
b=6356752.3\,\meter&:&\textnormal{Earth's polar radius (WGS84)}\\
\end{array}
\end{equation}

The E\"otv\"os effect represents two additional fictitious forces that arise during horizontal (north/east) motion on the surface of the Earth. The first term is the Coriolis force which can be interpreted as an apparent increase or decrease in Earth's centrifugal force due to east/west motion. The other term is known as the transport-rate effect, which arises due to the change in orientation of the vertical direction (along the ellipsoidal normal) as one moves across the Earth.

\subsection{GNSS Data Processing}

The GNSS observations were sampled at 1 Hz and processed using the Waypoint commercial software suite from Hexagon/NovAtel along with precise ephemerides from the Center for Orbit Determination in Europe (CODE). The software utilizes a Kalman filter to derive position estimates using a Precise Point Positioning (PPP) approach. These estimates can be translated from the GNSS antenna to the position of any other instrument on board the aircraft if the lever arm (instrument-antenna separation) is known. Assuming that the lever arm, $\mathbf{l}^b$, is specified along the front, right and down directions of the aircraft (body frame) and is positive in the direction from the instrument to the GNSS antenna, the position can be translated using the aircraft attitude as

\begin{equation}
\mathbf{p}_\text{sensor} = \mathbf{p}_\text{GNSS} - \mathbf{T} \, \mathbf{C}_b^n \, \mathbf{l}^b \; ,
\end{equation}
where $\mathbf{p} = \left[ \varphi , \lambda , h \right]$ denotes the position in geodetic coordinates, $\mathbf{C}_b^n$ is the body-to-navigation-frame transformation matrix and $\mathbf{T}$ is the Cartesian-to-curvilinear transformation matrix as

\begin{equation}
\mathbf{C}_b^n =
\left[ \begin{matrix}
\cos \beta \cos \gamma & {\footnotesize \begin{matrix} -\cos\alpha\sin\gamma \\ +\sin\alpha\sin\beta\cos\gamma \end{matrix} } & {\footnotesize \begin{matrix} \sin\alpha\sin\gamma \\ +\cos\alpha\sin\beta\cos\gamma \end{matrix} } \\
\cos \beta \sin \gamma & {\footnotesize \begin{matrix} \cos\alpha\cos\gamma \\ +\sin\alpha\sin\beta\sin\gamma \end{matrix} } & {\footnotesize \begin{matrix} -\sin\alpha\cos\gamma \\ +\cos\alpha\sin\beta\sin\gamma \end{matrix} } \\
-\sin\beta & \sin\alpha\cos\beta & \cos\alpha\cos\beta
\end{matrix} \right]
\end{equation}
where $\alpha$, $\beta$ and $\gamma$ denotes the bank, elevation and heading angle, respectively, and

\begin{equation}
\mathbf{T} =
\left[ \begin{matrix}
\frac{1}{M(\varphi)+h} & 0 & 0 \\
0 & \frac{1}{\left(N(\varphi)+h\right)\cos\varphi} & 0 \\
0 & 0 & 1
\end{matrix} \right] \; .
\end{equation}

The kinematic and E\"{o}tv\"{o}s accelerations in \eqref{eq: sensor equation} are derived from these translated GNSS position estimates using a central finite difference estimator.

\subsection{Gravity disturbance calculation}

The gravity disturbance is obtained by subtracting the GRS80 normal gravity model
taking into account altitude and latitude effects \cite{Torge2012}:

\begin{equation}  
g_0=\frac{a\cdot g_E\cdot \cos(\varphi)^2+b\cdot g_P\cdot \sin(\varphi)^2}{\sqrt{a^2\cdot \cos(\varphi)^2+b^2\cdot \sin(\varphi)^2}}
\cdot (1+\Gamma_1\cdot h\nonumber +\Gamma_2 \cdot h^2 ) 
\end{equation}
with :
\begin{equation}
\begin{array}{l}
g_E=9.78 032 677\;\text{m}\,\text{s}^{-2}\;\;\text{(GRS80)} \\       
g_P=9.83 218 637 \;\text{m}\,\text{s}^{-2}\;\;\text{(GRS80)}\\
\Gamma_1=-\frac{2}{a}\left(1+f+\frac{a^2\cdot b\cdot \omega_E^2}{GM}-2\cdot f \cdot \sin(\varphi)^2\right)\\
\Gamma_2=\frac{3}{a^2}\\
f=\frac{a-b}{a} \\
GM = 3.986005\cdot 10^{14}\; \text{m}^3\, \text{s}^{-3}\;\;\text{(GRS80)}\\
a=6378137.0\, \meter:\textnormal{\small Earth's equatorial radius (GRS80)}\\
b=6356752.3\,\meter:\textnormal{\small Earth's polar radius (GRS80)}\\
\end{array}
\end{equation}

\normalsize

This compute the ellipsoidal gravity in the GRS80 system.

\subsection{GIRAFE quantum gravimeter}\label{GIRAFE processing}

The data processing and the gravity estimation from the quantum gravimeter measurements and GNSS data is similar to the one used for our last airborne campaign \cite{Bidel2020}. The main difference comes from the low pass filter where a gaussian filter is used instead of a Bessel fourth-order filter. We also here do not correct alignment error of the platform. Indeed, during this campaign the estimated error alignment of the platform are small ($\leq$ 1.5 mrad) and the correction is not pertinent.

The Gaussian low pass filter used for the data processing is implemented in the Fourier domain. The FFT of the data are multiplied by the Gaussian function:
\begin{equation}
h(\omega)=\exp \left( -\frac{\omega^2 \sigma_t^2}{2}\right)
\end{equation}
In order to minimize border effects, the input signal of the filter is extended from both side with the symmetrically signal.
In the time domain, this is equivalent to convolute the data by a Gaussian function equal to
\begin{equation}
h(t)=\frac{1}{\sqrt{2\pi}\sigma_t}\exp \left(-\frac{t^2}{2\, \sigma_t^2}\right)
\end{equation}
$h(t)$ represents the response of the filter to a Dirac function. The filter is characterised by the Full-Width-Half-Maximum (FWHM) of this function which is equal to $\Delta \tau=\sqrt{8\,ln(2)}\sigma_t$. For a plane of velocity $v$, this gives a spatial resolution equal to $\Delta x=v\,\Delta \tau$.

The GNSS data and gravimeter data are first filter by a $\Delta t$=7 s low pass Gaussian filter. Then, the gravimeter data at 10 Hz are linearly interpolated on the time vector of GNSS data at 1 Hz. Before interpolation, the precise delay between GNSS and gravimeter data is adjusted to obtain the best correlation between acceleration measured from the gravimeter and acceleration derived from GNSS data. The delay is adjusted with a precision of 10 ms and is found to be constant for each flight. Then we select measurements acquired during plane straight trajectory  i.e. with a yaw rotation rate below 0.2 mrad/s. The measurements during change of direction do not allow precise gravity measurements and could perturb by border effect the gravity measurements in straight line. The gravimeter measurements are corrected for kinematic and E\"{o}tv\"{o}s acceleration. Finally, a Gaussian low pass filter is applied with a FWHM pulse response ranging from 25 s to 200 s.

\subsection{The L{\&}R Platform Gravimeter}
The L{\&}R sensor is mounted on a two-axis damped platform that keeps the sensitive axis of the gravimeter approximately aligned with the direction of the gravity vector, i.e. the plumb line. To derive gravity estimates, the specific force observations, $f$, are corrected for kinematic and E\"{o}tv\"{o}s accelerations according to \eqref{eq: sensor equation} and \eqref{eq: eotvos}. To transition from relative to absolute gravity estimates, a base reading, $g_\text{base}$, performed before the flight is subtracted from the observations and an external tie value, $g_\text{tie}$, is added to the observations as

\begin{equation}
g = f - \ddot{h} + a_{\text{E\"ot}} + \delta g _\text{tilt} + \left( g_\text{tie} - g_\text{base} \right) \; .
\end{equation}

In order to correct for any sensor misalignment, a tilt correction, $\delta g _\text{tilt}$, is introduced. A pair of accelerometers mounted on top of the sensor casing along the long- and cross-axes of motion are exploited to estimate the tilt angle of the sensor. Assuming that the long-axis accelerometer is tilted by an angle, $\phi_\text{long}$, the observed accelerations will represent a component of gravity, $g$, and a component of kinematic acceleration, $q_\text{long}$, projected onto the sensitive axis as

\begin{equation}
f_\text{long} = g \sin \phi_\text{long} + q_\text{long} \cos \phi_\text{long} \approx g \, \phi_\text{long} + q_\text{long} \; ,
\end{equation}
where small angle approximations are introduced. The kinematic acceleration, $q_\text{long}$, is derived from GNSS position estimates, projected onto the direction of travel and corrected for a horizontal E\"{o}tv\"{o}s effect. An estimate of the long-axis tilt angle is therefore

\begin{equation}
\phi_\text{long} \approx \frac{f_\text{long} - q_\text{long}}{g_0} \; ,
\end{equation}
where the standard value of gravity, $g_0 = 9.80665 \, \mathrm{m}/\mathrm{s}^2$, is used. Similar arguments hold for the cross-axis direction. From the estimated tilt angles, a tilt correction is formed as \cite[Eq.~2.11]{2002_Olesen}:

\begin{equation}
\delta g_\text{tilt} = \left( 1 - \cos \phi_\text{long} \cos \phi_\text{cross} \right) f + \sin \phi_\text{long} \, f_\text{long} + \sin \phi_\text{cross} \, f_\text{cross} \; ,
\end{equation}
which is derived under the assumption of small tilt angles. Following these corrections, the gravity estimates are filtered using a three-fold forward/backward zero-phase Butterworth filter with a (full-width) half power point of 134 s in the temporal domain. At an along-track velocity of 100 m/s, this corresponds to a FWHM spatial resolution of approximately 13.4 km. The relatively heavy filtering of the gravimeter observations were necessitated by the relatively turbulent flights, compared to earlier campaigns \cite{Forsberg2010}.

\subsection{The iMAR Strapdown Gravimeter}
The temperature stabilized IMU from iMAR Navigation was mounted in a strapdown configuration, meaning that the sensor casing is rigidly mounted to the chassis of the aircraft. The observations of the internal accelerometers and gyroscopes are sequentially integrated in a dead reckoning methodology to form independent estimates of attitude, velocity and heading. Because errors are integrated and will continue to increase using this approach, the navigation estimates are continuously corrected and the sensor biases calibrated in a Kalman filter framework by introducing GNSS position estimates. The state vector used in the Kalman filter is

\begin{equation}
\delta \mathbf{x} = \left[ \delta \boldsymbol{\psi} \, , \, \delta \mathbf{v} \, , \, \delta \mathbf{p} \, , \, \delta \mathbf{b}_g \, , \, \delta \mathbf{b}_a \, , \, \delta \triangle \mathbf{g} \, , \, \delta \triangle \dot{\mathbf{g}} \, , \, \delta \triangle \ddot{\mathbf{g}} \right] ^\top \; ,
\end{equation}
where $\boldsymbol{\psi}$, $\mathbf{v}$ and $\mathbf{p}$ denotes the attitude, velocity and position, respectively, $\mathbf{b}_g$ and $\mathbf{b}_a$ denotes the gyroscope and accelerometer biases, respectively, and $\triangle \mathbf{g}$ denotes the anomalous/disturbing gravity vector, with respect to the gravity model used to correct specific force observations during processing. In the above, dots denote derivative with respect to time and $\delta$ indicates an error-state implementation, meaning that the Kalman filter estimates errors on the inertial navigation solution, rather than full quantities.

The temporal evolution of the error-state vector, $\delta \mathbf{x}$, is described by a linear dynamic system model

\begin{equation}\label{eq: linear dynamic system model}
\delta \dot{\mathbf{x}} (t) = \mathbf{F} (t) \, \delta \mathbf{x} (t) + \mathbf{G} (t) \, \mathbf{w}_s (t) \; ,
\end{equation}
containing a white noise vector, $\mathbf{w}_s$, and system noise distribution matrix, $\mathbf{G}$, allowing the user to model sensor errors and their distribution onto the state variables. Additionally, the (re-)distribution of errors is determined by the motion-dependent system matrix

\begin{equation}
 \mathbf{F} =
 \left[
 \begin{array}{ccc|cc|ccc}
  \mathbf{F}_{11}^n & \mathbf{F}_{12}^n & \mathbf{F}_{13}^n & \boldsymbol{0}_3 & \mathbf{C}_ b^n & \boldsymbol{0}_3  & \boldsymbol{0}_3 & \boldsymbol{0}_3 \\
  \mathbf{F}_{21}^n & \mathbf{F}_{22}^n & \mathbf{F}_{23}^n & \mathbf{C}_ b^n & \boldsymbol{0}_3 & \mathbf{I}_3 & \boldsymbol{0}_3 & \boldsymbol{0}_3 \\
  \boldsymbol{0}_3  & \mathbf{F}_{32}^n & \mathbf{F}_{33}^n & \boldsymbol{0}_3 & \boldsymbol{0}_3 & \boldsymbol{0}_3  & \boldsymbol{0}_3 & \boldsymbol{0}_3 \\
  \hline
  \boldsymbol{0}_3  & \boldsymbol{0}_3  & \boldsymbol{0}_3  & \boldsymbol{0}_3 & \boldsymbol{0}_3 & \boldsymbol{0}_3  & \boldsymbol{0}_3 & \boldsymbol{0}_3 \\
  \boldsymbol{0}_3  & \boldsymbol{0}_3  & \boldsymbol{0}_3  & \boldsymbol{0}_3 & \boldsymbol{0}_3 & \boldsymbol{0}_3  & \boldsymbol{0}_3 & \boldsymbol{0}_3 \\
  \hline
  \boldsymbol{0}_3  & \boldsymbol{0}_3  & \boldsymbol{0}_3  & \boldsymbol{0}_3 & \boldsymbol{0}_3 & \boldsymbol{0}_3  & \mathbf{I}_3 & \boldsymbol{0}_3 \\
  \boldsymbol{0}_3  & \boldsymbol{0}_3  & \boldsymbol{0}_3  & \boldsymbol{0}_3 & \boldsymbol{0}_3 & \boldsymbol{0}_3  & \boldsymbol{0}_3 & \mathbf{I}_3 \\
  \boldsymbol{0}_3  & \boldsymbol{0}_3  & \boldsymbol{0}_3  & \boldsymbol{0}_3 & \boldsymbol{0}_3 & -\boldsymbol{\beta}^3  & -3\boldsymbol{\beta}^2 & -3\boldsymbol{\beta}
 \end{array}
 \right] \; ,
\end{equation}
with the elements of the upper left corner listed in \cite[Equations~14.64-14.71]{2013_Groves}, $\mathbf{C}_ b^n$ is the transformation matrix from the body- to navigation-frame and $\boldsymbol{\beta}$ is a $3 \times 3$ diagonal matrix containing temporal correlation parameters of the gravity anomaly vector.

The differential equation \eqref{eq: linear dynamic system model} is essentially solved for in each propagation interval between subsequent GNSS position estimates. This allows for the forward propagation of an error covariance matrix, $\mathbf{P}$, which is associated with the inertial navigation estimates. Since both the inertial and GNSS navigation estimates now have an associated error covariance matrix, these can be combined in a statistically optimal fashion by forming the Kalman filter gain

\begin{equation}
\mathbf{K} = \frac{\mathbf{P} \, \mathbf{H}^\top}{\mathbf{H} \, \mathbf{P} \, \mathbf{H}^\top + \mathbf{R}} \; ,
\end{equation}
where $\mathbf{R}$ is the error covariance of the GNSS position estimates and $\mathbf{H}$ is a measurement matrix, relating the state variables to the observations (position estimates). The error state vector, $\delta \mathbf{x}$, and error covariance matrix, $\mathbf{P}$, is then updated by forming the measurement innovation, $\delta \mathbf{z}$, as the difference between inertial and GNSS navigation estimates:

\begin{equation}\label{eq: position update}
\begin{aligned}
\delta \mathbf{x} & = \mathbf{K} \, \delta \mathbf{z}
= \mathbf{K} \, \left( \mathbf{p}_\text{GNSS} - \mathbf{H}
\left[ \begin{matrix} \boldsymbol{\psi}_\text{IMU} \\ \mathbf{v}_\text{IMU} \\ \mathbf{p}_\text{IMU} \\ \boldsymbol{0}_{15\times3} \end{matrix} \right]
\right) \\
\mathbf{P}^\text{updated} & = \mathbf{P} - \mathbf{K} \left( \mathbf{H} \, \mathbf{P} \right)\; .
\end{aligned}
\end{equation}

Once the errors on the state variables, $\delta \mathbf{x}$, are estimated, these are used to correct the inertial navigation estimates before continuing the dead reckoning navigation approach and to correct any sensor output for systematic errors, i.e. sensor bias. This is denoted as a closed-loop implementation of the Kalman filter and results in combined IMU/GNSS estimates of attitude, velocity, position and sensor biases. The attitude estimates are used in the computation of any E\"{o}tv\"{o}s and tilt effects, but the combined IMU/GNSS navigation solution should not be used to derive kinematic accelerations.

Since the error on the gravitational model used in processing is estimated in the Kalman filter, these error estimates can be added back to the model to arrive at gravity estimates. In this case, the temporal along-track evolution of the gravity disturbance is modelled as a third-order Gauss-Markov process, with a standard deviation, $\sigma_\text{GM3}$, and correlation parameter, $\beta_\text{GM3}$, which essentially controls the degree of smoothing the gravity estimates. These parameters are part of the system noise vector, $\mathbf{w}_s$, and system matrix, $\mathbf{F}(t)$, where the correlation matrix is formed as

\begin{equation}
 \boldsymbol{\beta} = 
 \left[ \begin{matrix}
  \beta_N & 0 & 0 \\
  0 & \beta_E & 0 \\
  0 & 0 & \beta_D
 \end{matrix} \right]
 = \left\vert v_\text{hor} \right\vert
 \left[ \begin{matrix}
  \beta_{N,\text{GM3}} & 0 & 0 \\
  0 & \beta_{E,\text{GM3}} & 0 \\
  0 & 0 & \beta_{D,\text{GM3}}
 \end{matrix} \right] \; ,
\end{equation}
using the along-track velocity, $\left\vert v_\text{hor} \right\vert^2 = v_N^2 + v_E^2$, to convert from spatial to temporal domains. The final estimates are derived by running the Kalman filter both forward and backward in time and combining these using a Rauch-Tung-Striebel smoother \cite[Chapter~5]{1974_Gelb}. In this approach, no further smoothing of the gravity estimates is needed.

\acknowledgments

The development of the ONERA quantum gravimeter GIRAFE was funded by ONERA, the French Defense Agency (DGA) and Shom. The airborne campaign was carried out with support from CNES (project n°3673, \cite{Bonvalot2018}) and ESA. We thank SAFIRE for their excellent support for the flights and the installation of the complex system in the ATR-42. The data used in this manuscript are available from the corresponding author upon request.

%% ------------------------------------------------------------------------ %%
%% References and Citations

%%%%%%%%%%%%%%%%%%%%%%%%%%%%%%%%%%%%%%%%%%%%%%%
%
% \bibliography{<name of your .bib file>} don't specify the file extension
%
% don't specify bibliographystyle
%%%%%%%%%%%%%%%%%%%%%%%%%%%%%%%%%%%%%%%%%%%%%%%

%\bibliography{Bibliographie_v3}

%Reference citation instructions and examples:
%
% Please use ONLY \cite and \citeA for reference citations.
% \cite for parenthetical references
% ...as shown in recent studies (Simpson et al., 2019)
% \citeA for in-text citations
% ...Simpson et al. (2019) have shown...
%
%
%...as shown by \citeA{jskilby}.
%...as shown by \citeA{lewin76}, \citeA{carson86}, \citeA{bartoldy02}, and \citeA{rinaldi03}.
%...has been shown \cite{jskilbye}.
%...has been shown \cite{lewin76,carson86,bartoldy02,rinaldi03}.
%... \cite <i.e.>[]{lewin76,carson86,bartoldy02,rinaldi03}.
%...has been shown by \cite <e.g.,>[and others]{lewin76}.
%
% apacite uses < > for prenotes and [ ] for postnotes
% DO NOT use other cite commands (e.g., \citet, \citep, \citeyear, \nocite, \citealp, etc.).
%

\end{document}